\documentclass[journal]{IEEEtran}
\usepackage{verbatim}
\usepackage{subfiles}
\usepackage{cite}
\usepackage{graphicx}
\usepackage{amsmath}
\usepackage{lipsum}
\usepackage{changepage}
\usepackage{float}
\usepackage{algorithm}
\usepackage{algpseudocode}
\usepackage{booktabs}
\usepackage{multirow}
\usepackage{nomencl}
\usepackage{enumitem}
\usepackage{xcolor}

\usepackage[most]{tcolorbox}
\usepackage{amsmath,amssymb,bm}

\usepackage{subcaption}
\usepackage{url}

\makenomenclature
\usepackage{etoolbox}
\usepackage{upgreek}
\usepackage{mathtools}

\usepackage{ifthen}
 \renewcommand{\nomgroup}[1]{%
  \item[\bfseries
  \ifthenelse{\equal{#1}{A}}{Parameters}{%
  \ifthenelse{\equal{#1}{B}}{Variables}{%
  \ifthenelse{\equal{#1}{O}}{Indices and sets}{}}}%
  ]}

\ifCLASSINFOpdf
\else
\fi

\usepackage{color}
\usepackage{soul}
\soulregister\cite7
\soulregister\ref7

\usepackage[dvipsnames]{xcolor}

\begin{document}

\title{Source Side Mitigation of AI Datacenter Power Fluctuations with a Hybrid Energy Storage System and Residual Differentiable Predictive Control}

\author{Haiyang You, Chengwei Lou, and Jin Yang%
\thanks{H.~You and J.~Yang are with the James Watt School of Engineering,
University of Glasgow, Glasgow G12 8QQ, UK\@.
C.~Lou is with the College of Information and Electrical Engineering,
China Agricultural University, Beijing 100083, China.}}

\markboth{IEEE TRANSACTIONS ON [JOURNAL NAME]}%
{H. YOU et al.: SOURCE SIDE MITIGATION OF AI DATACENTER POWER FLUCTUATIONS}
%



\maketitle

\begin{abstract}
The rapid growth of hyperscale AI datacenters introduces structured,
workload-driven active power fluctuations at the point of interconnection.
These fluctuations appear to the grid as time-varying disturbance
injections that cannot be captured by conventional peak or average load
representations. To reduce the residual power disturbance before it
propagates into the power system, this paper proposes a hybrid
energy storage system with differentiable predictive control (HESS-DPC)
framework for datacenter-side power smoothing. A workload-driven
disturbance model is first established, representing the
point of interconnection load deviation as the superposition of
training and fine-tuning workloads to capture the structured forcing
inputs that can excite frequency dynamics. A frequency
decomposition rule-based controller then allocates this deviation between a battery
energy storage system (BESS) and a supercapacitor (SC), assigning the
energy-dominant component to the BESS and the fast-varying component to
the SC. To overcome the anticipation and constraint limitations of fixed
frequency decomposition, a residual differentiable predictive control
policy is trained offline to compute finite-horizon command corrections
around the rule-based baseline while enforcing a one-step safeguard.
Simulations on the NPCC 140-bus system show that HESS-DPC reduces
grid-side residual deviations during workload transitions, improves SC
state of charge sustainability over extended operation, and reduces
generator peak-to-peak frequency deviations by more than 80\% across all
monitored generators, with the worst affected generator response falling
from 15.1~mHz to 1.3~mHz. These results confirm that local
active power smoothing at the datacenter point of interconnection
can substantially mitigate frequency disturbances caused by AI workloads.
\end{abstract}


%
\IEEEpeerreviewmaketitle

\linespread{0.81}\selectfont

{

\small


\makenomenclature
\renewcommand{\nomgroup}[1]{} %

\nomenclature[N01]{$\mathcal{N}$}{Set of all buses}
\nomenclature[N02]{$\mathcal{L}$}{Set of all transmission lines}
\nomenclature[N03]{$\mathcal{G}$}{Set of all generators}
\nomenclature[N04]{$\mathcal{G}_n$}{Set of generators connected to bus $n$}
\nomenclature[N05]{$\mathcal{D}_n$}{Set of demand units at bus $n$}
\nomenclature[N06]{$\mathcal{L}_n^{+}$}{Set of lines with bus $n$ as sending end}
\nomenclature[N07]{$\mathcal{L}_n^{-}$}{Set of lines with bus $n$ as receiving end}
\nomenclature[N08]{$B_l$}{Susceptance of line $l$}
\nomenclature[N08]{$n(l),\, m(l)$}{Sending/receiving buses of line $l$}
\nomenclature[N09]{$g,\,j,\,l,\,n,\,m$}{Indices for generators, data centers, lines, and buses}

\nomenclature[N10]{$\alpha,\,\beta$}{Coefficients of the GPU energy-consumption model}
\nomenclature[N11]{$\mathrm{PUE}_j$}{Power usage effectiveness of data center $j$}
\nomenclature[N12]{$a$}{Weighting coefficient between the physical and neural models}
\nomenclature[N13]{$c_g$}{Carbon emission factor of generator $g$}
\nomenclature[N14]{$F_l$}{Capacity limit of transmission line $l$}
\nomenclature[N15]{$P_{\mathrm{idle}}$}{Idle power consumption of a GPU}
\nomenclature[N16]{$\lambda$}{Weighting factor for the constraint penalty}
\nomenclature[N20]{$\xi$}{Parameters of the neural network}

\nomenclature[N21]{$\theta_n$}{Voltage angle at bus $n$}
\nomenclature[N22]{$f_l$}{Power flow on transmission line $l$}
\nomenclature[N23]{$p_g$}{Power output of generator $g$}
\nomenclature[N24]{$d_n$}{Base demand at bus $n$ (used in DC-OPF balance)}
\nomenclature[N25]{$P_n^{\mathrm{dem}}$}{Total power demand at bus $n$ (includes DC load)}
\nomenclature[N26]{$P_n^{\mathrm{gen}}$}{Total power generation at bus $n$}
\nomenclature[N27]{$E_n^{\mathrm{gen}}$}{Generation-related carbon emissions at bus $n$}
\nomenclature[N28]{$P_{\mathrm{DC},j}$}{Power consumption of data center $j$}
\nomenclature[N29]{$P_{\mathrm{GPU}}(t)$}{Total GPU power consumption at time $t$}
\nomenclature[N30]{$u(t)$}{GPU utilization rate at time $t$}
\nomenclature[N31]{$n_{q,j}$}{Number of AI queries at data center $j$}
\nomenclature[N32]{$P_{\mathrm{neural}}(t)$}{Neural model power component at time $t$}
\nomenclature[N33]{$P_{\mathrm{physical}}(t)$}{Physical model power component at time $t$}
\nomenclature[N34]{$P_{\mathrm{obs}}(t)$}{Observed GPU power at time $t$}
\nomenclature[N37]{$CE$}{Total system carbon emissions}
\nomenclature[N39]{$\mathrm{LMCE}$}{Locational marginal carbon emissions}
\nomenclature[N40]{$\mathrm{ACE}$}{Average carbon emissions}

\printnomenclature[2cm]

}

\section{Introduction}
The rapid expansion of hyperscale AI datacenters is introducing new grid-integration challenges for large loads in power systems. In conventional planning and adequacy studies, a large customer is often characterized by aggregate demand descriptors, such as peak demand or representative load profiles. These abstractions are useful for steady-state capacity assessment, but they may fail to capture short-term active power variations and fast ramping behaviors that enter the grid as dynamic disturbances~\cite{nerc2025largeloads,ross2025emt,shehabi2024datacenter}. For AI datacenters, the grid impact is therefore determined not only by the connected power level, but also by the magnitude, ramp rate, and timescale of the resulting power variations.

A key source of these variations is the workload execution dynamics of power-intensive AI computing. During large-scale training jobs, many GPUs repeatedly alternate between computation-heavy phases and lower-utilization communication or synchronization phases~\cite{choukse2025power}. Fine-tuning tasks and smaller concurrent jobs introduce additional variations at different power levels and timescales~\cite{ye2024deep,patel2024characterizing}. When workload phases start, end, or transition, the number of devices operating at high utilization changes accordingly, producing active power fluctuations at the facility level~\cite{wilkins2026servers}. These fluctuations can create ramping stress and persistent disturbances at the point of interconnection, increasing the dynamic stress associated with large-load integration~\cite{OKeefe2025EventRecords,ParkerSter}.

Prior studies have examined datacenter power variability at both the facility and grid levels. Measurement-based studies have characterized it at the facility level by using GPU-level power traces to construct aggregate demand profiles~\cite{li2024unseen,jimenez2025data,borghesi2023m100,radovanovic2021power}. Beyond facility-scale characterization, recent reliability-oriented studies indicate that the tightly synchronized, periodic compute cycles of AI training datacenters can induce sustained load oscillations that propagate into the bulk power system across a wide frequency range~\cite{naspi2026oscillations}. Because such fast, structured variations are not captured by conventional peak or average load descriptors, dedicated fast-timescale and electromagnetic-transient modeling of datacenter loads has been advocated for grid-level studies~\cite{nrel2026framework,ross2025emt}. From a modeling perspective, recent works represent AI datacenter loads as structured forcing inputs in power-system dynamic studies and show that these variations can sustain oscillatory responses rather than only produce isolated transient events~\cite{ko2026wide}. Ko et al.~\cite{ko2025mitigation} further investigate the use of hybrid energy storage to reduce datacenter-induced ramping and fluctuation stresses under prolonged stochastic training cycles. These studies confirm that datacenter power variability can have direct dynamic impacts on the bulk power system. However, they primarily characterize or model the disturbance, and even where hybrid storage is considered for ramping and fluctuation reduction~\cite{ko2025mitigation}, source-side suppression of residual active-power disturbances at the datacenter point of interconnection explicitly addressed under time-varying workload conditions remains insufficiently studied.

The practical importance of this source-side mitigation gap is reinforced by recent reliability guidance from the North American Electric Reliability Corporation (NERC) on emerging large loads. NERC notes that some large loads, including computational loads, can exhibit second-to-second and minute-to-minute power oscillations together with fast ramping capability, which complicates short-term operational forecasting for system operators. Such fast load variations can produce demand swings within seconds or minutes that stress balancing reserves and frequency-control resources, while abrupt load changes or disconnections may cause system imbalance and frequency instability. To address these risks, NERC recommends incorporating large-load variability into balancing assessments so that sufficient regulating capability and coordination can be maintained under existing control-performance and ACE-limit requirements. For AI training datacenters, it further identifies mitigation options such as software mitigation, GPU power smoothing, and rack-level energy storage, and points to interconnection requirements based on oscillation attenuation metrics, real-power amplitude variation thresholds, and amplitude-frequency limits for oscillatory demand~\cite{nerc2026emerginglargeloads}. These recommendations directly motivate a point-of-interconnection-level smoothing mechanism that suppresses residual active-power disturbances at the datacenter-side, thereby reducing the burden imposed on grid-side balancing and frequency-control functions.

While NERC identifies device-level options such as GPU power smoothing and rack-level storage, this work focuses on a hybrid energy storage system (HESS) installed at the datacenter boundary. This boundary-level interface can operate autonomously at the facility level without requiring transmission-level coordination or upstream control modifications~\cite{ross2025emt}. The hybrid form is motivated by the spectral characteristics of the datacenter power disturbance. Because the disturbance contains both a slow, energy-dominant component and a fast, power-dominant component~\cite{Ko2026Oscillations}, a single storage technology is not well suited to handle both within practical operating limits~\cite{choukse2025power}. Battery energy storage systems (BESSs) provide a larger energy buffer but are less suitable for sustained high-frequency tracking~\cite{nerc2023gfm}, whereas supercapacitors (SCs) can respond quickly to transient deviations but cannot sustain long-duration compensation due to their limited energy capacity~\cite{xiao2015hierarchical}. A HESS combines these complementary properties and provides a structured local interface for reducing AI datacenter power fluctuations at the source.

\begin{figure}[t]
\centering
\includegraphics[width=\linewidth]{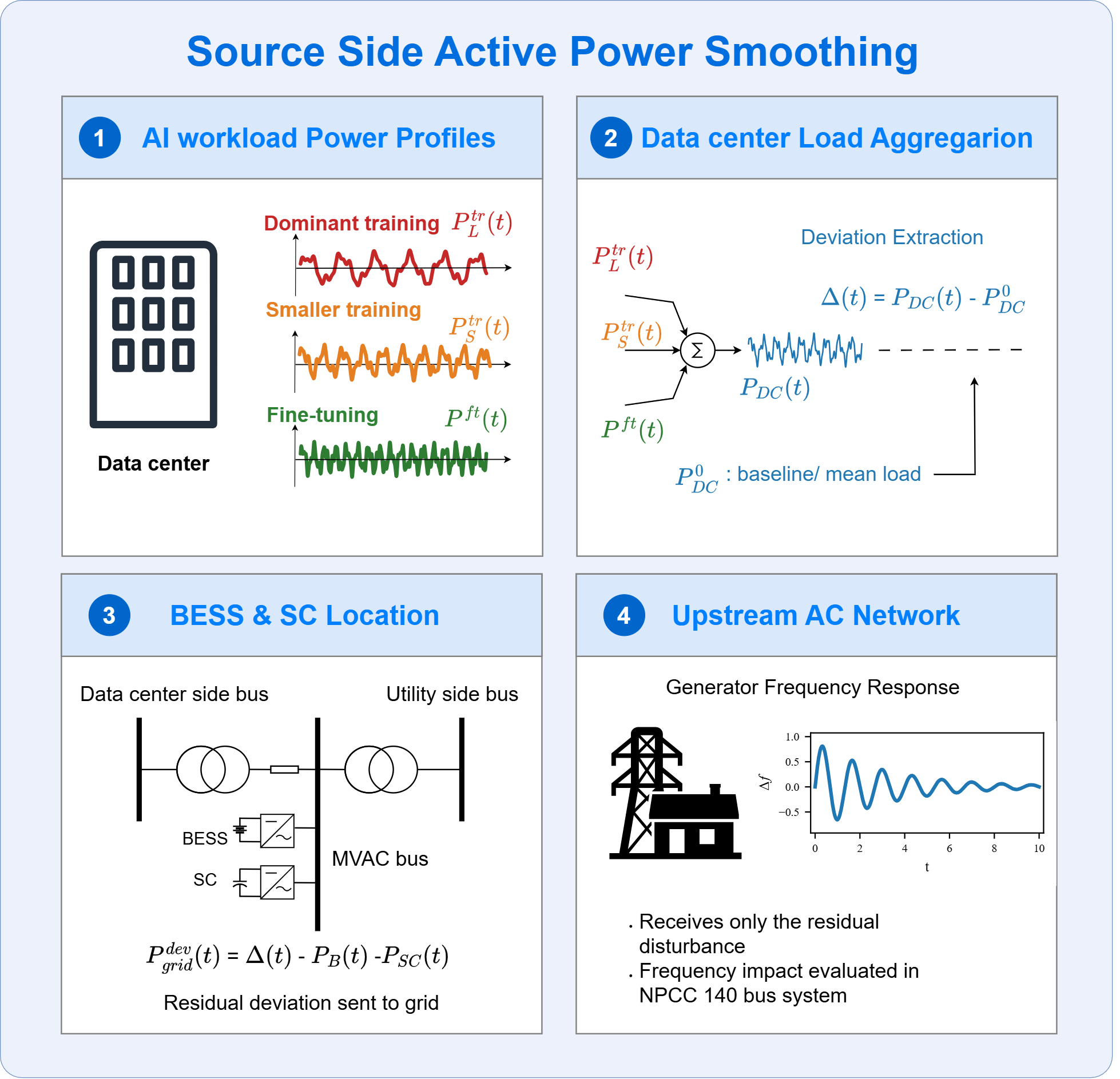}
\caption{Source-side active power smoothing framework for AI datacenter fluctuations.}
\label{fig:frame}
\end{figure}

Rule-based HESS controllers provide a natural first step for this smoothing problem~\cite{shim2020virtual}. A frequency-decomposition controller can assign the fast component of the datacenter disturbance to the SC and the remaining component to the BESS~\cite{chmielewski2021modified}. Such a structure is physically meaningful and easy to implement, but its filter settings and allocation parameters are typically selected for a prescribed operating condition. When the workload profile changes in amplitude, frequency content, or ramping pattern, the same fixed controller may leave non-negligible residual deviations, especially during workload transition intervals~\cite{cao2011new}. Moreover, fixed rule-based allocation does not explicitly anticipate future disturbances or jointly optimize residual power reduction and device operating constraints over a prediction horizon.

These limitations motivate the incorporation of a predictive refinement layer. DPC offers such a mechanism without requiring an online optimization problem to be solved at every sampling instant~\cite{amos2018differentiable}. It combines a known prediction model with an offline training procedure based on a model-predictive-control (MPC)-inspired finite-horizon objective~\cite{drgovna2022differentiable}. In this work, DPC is adopted in residual form: rather than generating the complete BESS and SC commands directly, it computes finite-horizon corrections around the rule-based HESS commands. This design preserves the physical BESS/SC allocation structure, limits the scope of the correction task, and improves performance in operating regions where the fixed baseline is most limited~\cite{silver2018residual,johannink2019residual}. The training objective penalizes grid-side residual power, command variation, and violations of power, ramp-rate, and state-of-charge limits~\cite{chen2018approximating,mayne2000constrained}. The proposed HESS-DPC framework first extracts the workload-driven active power deviation, applies frequency-based BESS/SC allocation, and then uses residual DPC to reduce the grid-side residual. The resulting disturbance is injected into the Northeast Power Coordinating Council (NPCC) 140-bus system to quantify the reduction in workload-driven generator frequency deviations. The overall source-side mitigation pathway is illustrated in Fig.~\ref{fig:frame}. The main contributions of this paper are summarized as follows.

\begin{itemize}

\item \textbf{A source-side disturbance model is developed to characterize the active power forcing input imposed by AI datacenter workloads.} The datacenter active power deviation from its mean value is modeled as the disturbance input and constructed from dominant training, smaller training, and fine-tuning workloads, retaining the structured fluctuation components relevant to generator frequency dynamics.

\item \textbf{A frequency-decomposition HESS interface is established for point-of-interconnection smoothing.} The workload-driven disturbance is separated into energy-dominant and fast-varying components and allocated to the BESS and SC, respectively, under power, ramp-rate, and state-of-charge constraints.

\item \textbf{A residual DPC strategy is proposed to compensate for the limitations of fixed rule-based HESS allocation.} The policy learns finite-horizon corrections around the rule-based BESS/SC commands through a differentiable HESS rollout with an MPC-inspired loss, improving smoothing during workload transition intervals while preserving the physical baseline allocation.

\item \textbf{The grid-level impact of source-side smoothing is evaluated in the NPCC 140-bus system.} The residual disturbance after HESS-DPC compensation is injected into the bulk-system model to quantify the attenuation of workload-driven generator frequency deviations.

\end{itemize}


The remainder of this paper is organized as
follows. Section~\ref{sec:system_modeling} presents the datacenter load
model and the HESS model. Section~\ref{sec:dpc_smoothing} develops the
rule-based baseline controller and the residual DPC method.
Section~\ref{sec:case_studies} presents the NPCC 140-bus simulation
studies, and Section~\ref{sec:conclusion} concludes the paper.
\section{Datacenter Level Load Aggregation}
\label{sec:system_modeling}
The active power demand at the datacenter point of interconnection is driven by the aggregate of concurrently executing AI workloads. These workloads exhibit periodic phase structures whose superposition produces sustained fluctuations at the facility level. Following the stochastic workload model in~\cite{Ko2026Oscillations}, which characterizes the power level statistics of training and fine-tuning workloads from GPU profiling measurements, the considered operating condition consists of one dominant large-scale training workload, one smaller training workload, and one fine-tuning workload. The aggregate computational load profile is written as

\begin{equation}
P_{\Sigma}(t)
=
P^{\mathrm{tr}}_{L}(t)
+
P^{\mathrm{tr}}_{S}(t)
+
P^{\mathrm{ft}}(t),
\label{eq:aggregate_compute_load}
\end{equation}
where $P^{\mathrm{tr}}_{L}$, $P^{\mathrm{tr}}_{S}$, and $P^{\mathrm{ft}}$ denote the dominant training, smaller training, and fine-tuning profiles, respectively. The smaller workloads are scaled as $P^{\mathrm{tr}}_{S0}=\kappa_S P^{\mathrm{tr}}_{L0}$ and $P^{\mathrm{ft}}_{0}=\kappa_F P^{\mathrm{tr}}_{L0}$. The aggregate is then scaled to the target datacenter operating level:
\begin{equation}
P_{\mathrm{DC}}(t)
=
\gamma_{\mathrm{DC}} P_{\Sigma}(t),
\label{eq:datacenter_load}
\end{equation}
where $\gamma_{\mathrm{DC}}>0$ is a case study scaling factor fixed before control design. This factor uniformly scales the aggregate workload profile to the target datacenter operating level while preserving the relative temporal structure of the underlying training and fine-tuning fluctuations. Thus, $P_{\mathrm{DC}}(t)$ represents the equivalent active power demand of the datacenter as seen from the grid side. This profile is used as the load input for the subsequent HESS smoothing control.

The HESS is designed to compensate the fluctuation component of the datacenter load rather than its average demand. In the simulation study, the baseline datacenter power is computed offline over the evaluation horizon as
\begin{equation}
P^{0}_{\mathrm{DC}}
=
\frac{1}{T}
\int_{0}^{T}
P_{\mathrm{DC}}(t)\,dt,
\label{eq:mean_dc_power_continuous}
\end{equation}
where $T$ is the simulation horizon. For a discrete time profile, this quantity is computed as
\begin{equation}
P^{0}_{\mathrm{DC}}
=
\frac{1}{N}
\sum_{n=0}^{N-1}
P_{\mathrm{DC}}[n],
\label{eq:mean_dc_power_discrete}
\end{equation}
where $N$ is the number of samples. In practical online operation, this baseline may be replaced by a scheduled operating reference or a slowly updated moving average estimate, while the controller continues to act on the corresponding deviation signal.

The datacenter power deviation is then defined by
\begin{equation}
\Delta(t)
=
P_{\mathrm{DC}}(t)-P^{0}_{\mathrm{DC}},
\label{eq:dc_deviation_continuous}
\end{equation}
or, in discrete time,
\begin{equation}
\Delta[n]
=
P_{\mathrm{DC}}[n]-P^{0}_{\mathrm{DC}}.
\label{eq:dc_deviation_discrete}
\end{equation}
A positive deviation $\Delta(t)>0$ indicates that the datacenter demand exceeds its baseline level, whereas a negative deviation $\Delta(t)<0$ indicates that the demand is below the baseline level. By construction, both $\Delta(t)$ and $\Delta[n]$ have zero mean over the evaluation horizon, so the HESS is not required to supply net energy. They are used as the disturbance inputs for the HESS smoothing problem.

\subsection{HESS Modeling and Reference Power Allocation}

To mitigate the datacenter power deviation seen from the grid side, a hybrid energy storage system is installed at the point of interconnection. The HESS consists of a battery energy storage system (BESS) and a supercapacitor (SC). The two devices have complementary characteristics: the BESS offers a larger energy buffer suited to slower variations, while the SC provides a faster transient response for rapidly varying components.

Throughout the mathematical formulation, the subscript $\mathrm{B}$ denotes the BESS branch, while the subscript $\mathrm{SC}$ denotes the supercapacitor branch.

Following the discharged power positive convention, the remaining power deviation supplied by the grid is expressed as
\begin{equation}
P^{\mathrm{dev}}_{\mathrm{grid}}(t)
=
\Delta(t)
-
P_{\mathrm{B}}(t)
-
P_{\mathrm{SC}}(t),
\label{eq:grid_residual_continuous}
\end{equation}
where $P_{\mathrm{B}}(t)$ and $P_{\mathrm{SC}}(t)$ are the actual output powers of the BESS and SC, respectively. Positive storage power denotes discharging power delivered to compensate an above-baseline demand deviation, while negative storage power denotes charging associated with a below-baseline deviation. Perfect compensation corresponds to $P_{\mathrm{B}}(t)+P_{\mathrm{SC}}(t)=\Delta(t)$, in which case the grid-side residual in~\eqref{eq:grid_residual_continuous} vanishes; the smoothing objective is to bring the realized output pair as close to this ideal as the device constraints permit.

Each storage unit is represented by a first-order power response model. For $d\in\{\mathrm{B},\mathrm{SC}\}$, the transfer function from the control input $U_d(s)$ to the actual output power $P_d(s)$ is
\begin{equation}
G_d(s)
=
\frac{P_d(s)}{U_d(s)}
=
\frac{1}{\tau_d s+1},
\label{eq:first_order_device_model}
\end{equation}
where $\tau_d$ is the response time constant. The complementary characteristics of the two devices are summarized by
\begin{equation}
E^{\mathrm{cap}}_{\mathrm{B}}
\gg
E^{\mathrm{cap}}_{\mathrm{SC}},
\qquad
R^{\max}_{\mathrm{B}}
<
R^{\max}_{\mathrm{SC}},
\qquad
\tau_{\mathrm{B}}
\gg
\tau_{\mathrm{SC}}.
\label{eq:hess_characteristics}
\end{equation}

The storage units are required to operate within prescribed power, ramp rate, and state of charge limits:
\begin{align}
|P_d(t)| &\leq P^{\max}_d, \label{eq:power_constraint}\\
\left|\frac{dP_d(t)}{dt}\right| &\leq R^{\max}_d, \label{eq:ramp_constraint}
\end{align}
\begin{equation}
\mathrm{SoC}^{\min}_d
\leq
\mathrm{SoC}_d(t)
\leq
\mathrm{SoC}^{\max}_d,
\qquad
d\in\{\mathrm{B},\mathrm{SC}\}.
\label{eq:soc_constraint}
\end{equation}
These relations define the operating requirements of the HESS smoothing problem. Their specific treatment within the discrete time predictive control formulation is introduced in Section~\ref{sec:dpc_smoothing}.

The reference powers are generated from $\Delta(t)$ through a frequency-decomposition allocation.

A first-order low-pass filter extracts the slow component:
\begin{equation}
\Delta_{\mathrm{LPF}}(s)
=
G_{\mathrm{LPF}}(s)\Delta(s),
\label{eq:lpf_output}
\end{equation}
where
\begin{equation}
G_{\mathrm{LPF}}(s)
=
\frac{2\pi f_{\mathrm{split}}}
{s+2\pi f_{\mathrm{split}}}.
\label{eq:lpf_tf}
\end{equation}
The complementary fast component is
\begin{equation}
\Delta_{\mathrm{HPF}}(s)
=
\Delta(s)
-
\Delta_{\mathrm{LPF}}(s).
\label{eq:hpf_component}
\end{equation}

Before being assigned to the SC, the fast component is further conditioned by a first-order shaping filter:
\begin{equation}
\Delta_{\mathrm{SC}}(s)
=
G_{\mathrm{shape}}(s)
\Delta_{\mathrm{HPF}}(s),
\label{eq:sc_shaped_component}
\end{equation}
with
\begin{equation}
G_{\mathrm{shape}}(s)
=
\frac{2\pi f_{\mathrm{shape}}}
{s+2\pi f_{\mathrm{shape}}}.
\label{eq:shape_tf}
\end{equation}
This shaping stage limits excessively sharp high-frequency content in the SC channel while preserving its role in compensating fast disturbance variations. In the time domain, the SC is assigned $\Delta_{\mathrm{SC}}(t)$ and the BESS compensates the remainder $\Delta(t)-\Delta_{\mathrm{SC}}(t)$, so that their sum equals $\Delta(t)$ before projection and device dynamics. Note that $\Delta(t)-\Delta_{\mathrm{SC}}(t)$ retains a residual fast component because $G_{\mathrm{shape}}(s)$ provides approximate rather than perfect frequency separation; in practice, this residual has limited influence on the realized BESS output because the larger time constant $\tau_{\mathrm{B}}$ attenuates high-frequency content in $P_{\mathrm{B}}(t)$. The discrete-time realization, including ZOH discretization and operating-limit treatment, is presented in Section~\ref{sec:dpc_smoothing}.

\section{Power Smoothing via Differentiable Predictive Control}
\label{sec:dpc_smoothing}

The discrete-time grid-side residual is
\begin{equation}
P^{\mathrm{dev}}_{\mathrm{grid}}[k]
=
\Delta[k]
-
P_{\mathrm{B}}[k]
-
P_{\mathrm{SC}}[k],
\label{eq:grid_residual_discrete}
\end{equation}
where $P_{\mathrm{B}}[k]$ and $P_{\mathrm{SC}}[k]$ are the actual BESS and SC powers. Reducing $|P^{\mathrm{dev}}_{\mathrm{grid}}[k]|$ means that a larger portion of the datacenter fluctuation is compensated locally by the HESS.

A two layer control structure is developed. The first layer is a rule-based HESS baseline controller that allocates the disturbance according to the complementary response characteristics of the BESS and SC. The second layer is a residual differentiable predictive control (DPC) policy that generates finite-horizon command corrections around the baseline trajectory. The resulting design combines physically interpretable frequency-based allocation with predictive refinement over a short look ahead window.

\subsection{Rule-Based HESS Baseline}
\label{sec:baseline_controller}

The baseline controller allocates $\Delta[k]$ between the BESS and SC via frequency decomposition. A discrete-time low-pass filter, obtained by ZOH discretization of~\eqref{eq:lpf_tf} with sampling period $T_s$, extracts the slow component:
\begin{equation}
\Delta_{\mathrm{LPF}}[k]
=
\alpha_{\mathrm{LPF}}
\Delta_{\mathrm{LPF}}[k-1]
+
(1-\alpha_{\mathrm{LPF}})
\Delta[k],
\label{eq:lpf_discrete}
\end{equation}
with
\begin{equation}
\alpha_{\mathrm{LPF}}
=
e^{-2\pi f_{\mathrm{split}}T_s}.
\label{eq:lpf_alpha}
\end{equation}
The corresponding fast component is
\begin{equation}
\Delta_{\mathrm{HPF}}[k]
=
\Delta[k]
-
\Delta_{\mathrm{LPF}}[k].
\label{eq:hpf_discrete}
\end{equation}

Before becoming the SC reference, the fast component is shaped by another first-order filter:
\begin{equation}
\Delta_{\mathrm{SC}}[k]
=
\alpha_{\mathrm{shape}}
\Delta_{\mathrm{SC}}[k-1]
+
(1-\alpha_{\mathrm{shape}})
\Delta_{\mathrm{HPF}}[k],
\label{eq:shape_discrete}
\end{equation}
where
\begin{equation}
\alpha_{\mathrm{shape}}
=
e^{-2\pi f_{\mathrm{shape}}T_s}.
\label{eq:shape_alpha}
\end{equation}
The cutoff frequency $f_{\mathrm{shape}}$ determines the bandwidth of the shaped SC reference.
Throughout this section, $\Pi_d(\cdot)$ denotes scalar projection onto the instantaneous power command interval defined by the device power rating:
\begin{align}
\Pi_d(z)
=
\min\!\left\{
P^{\max}_d,\,
\max\!\left\{
-P^{\max}_d,\,
z
\right\}
\right\},
\quad d\in\{\mathrm{B},\mathrm{SC}\}.
\label{eq:projection_operator}
\end{align}

This projection enforces the instantaneous command magnitude bounds, while realized power output, ramp rate, and SoC requirements are incorporated into the predictive objective through dedicated penalty terms.

The baseline SC reference is obtained by projecting the shaped fast component:
\begin{equation}
P^{\mathrm{ref},0}_{\mathrm{SC}}[k]
=
\Pi_{\mathrm{SC}}
\bigl(
\Delta_{\mathrm{SC}}[k]
\bigr).
\label{eq:sc_projected_ref}
\end{equation}
The remaining disturbance is assigned to the BESS and projected onto its instantaneous admissible range:
\begin{equation}
P^{\mathrm{ref},0}_{\mathrm{B}}[k]
=
\Pi_{\mathrm{B}}
\left(
\Delta[k]
-
P^{\mathrm{ref},0}_{\mathrm{SC}}[k]
\right).
\label{eq:bess_projected_ref}
\end{equation}

When neither projection is active, the baseline references preserve the exact balance relation inherited from the unconstrained allocation. When projection becomes active, the reference pair remains bounded within the prescribed instantaneous command limits, while the unallocated portion appears naturally in the residual deviation in~\eqref{eq:grid_residual_discrete}. This treatment keeps the baseline allocation consistent with the constrained smoothing objective.

To obtain an explicit and easily previewed baseline command sequence fully aligned with the predictive rollout, the local command mapping is defined as
\begin{equation}
U^{0}_{d}[k]
=
P^{\mathrm{ref},0}_{d}[k],
\qquad
d\in\{\mathrm{B},\mathrm{SC}\}.
\label{eq:baseline_static_mapping}
\end{equation}

By propagating the LPF~\eqref{eq:lpf_discrete} and shaping filter~\eqref{eq:shape_discrete} over the previewed disturbance $\Delta_{k:k+N_p-1}$ from the current filter state, the baseline controller generates the command sequence
\begin{equation}
\mathcal{T}^{0}_{k}
=
\bigl\{
U^{0}_{\mathrm{B},0:N_p-1},\,
U^{0}_{\mathrm{SC},0:N_p-1}
\bigr\},
\label{eq:baseline_trajectory}
\end{equation}
where the subscripts $0:N_p-1$ denote local prediction step indices within the window initialized at control instant $k$ (globally $k:k+N_p-1$). This command sequence serves as the reference around which the residual DPC correction is constructed.

\subsection{Residual DPC Policy}
\label{sec:residual_dpc}

DPC uses a differentiable finite-horizon prediction model to optimize a closed-loop policy offline. The policy is written in residual form: instead of generating complete BESS and SC commands directly, it outputs bounded command corrections that refine the rule-based baseline.

At time step $k$, the policy maps a feature vector $\xi_k$ to a sequence of residual corrections over the prediction horizon $N_p$:
\begin{equation}
\delta U_{0:N_p-1}
=
\pi_{\Theta}(\xi_k),
\label{eq:dpc_policy_output}
\end{equation}
where $\Theta$ denotes the policy parameters. For prediction step $j$,
\begin{equation}
\delta U_j
=
\begin{bmatrix}
\delta U_{\mathrm{B},j}\\
\delta U_{\mathrm{SC},j}
\end{bmatrix},
\qquad
j=0,\ldots,N_p-1.
\label{eq:residual_action_vector}
\end{equation}

The command used in the prediction rollout is
\begin{equation}
U_{d,j}
=
\Pi_d
\left(
U^{0}_{d,j}
+
\delta U_{d,j}
\right),
\qquad
d\in\{\mathrm{B},\mathrm{SC}\},
\label{eq:projected_dpc_command}
\end{equation}
where $U^{0}_{d,j}$ is the baseline command at prediction step $j$ within the current window.

Each residual component is bounded by
\begin{equation}
\delta U_{d,j}
=
\delta U^{\max}_d
\tanh(\hat{u}_{d,j}),
\qquad
d\in\{\mathrm{B},\mathrm{SC}\},
\label{eq:bounded_residual_output}
\end{equation}
with $\hat{u}_{d,j}$ the raw network output and $\delta U^{\max}_d$ the prescribed correction limit.

The feature vector includes the recent disturbance history, a short horizon disturbance preview, the current HESS state, and the baseline command preview:
\begin{multline}
\xi_k
=
\biggl[
\frac{\Delta_{k-N_h+1:k}}{\sigma_{\Delta}},\;
\frac{\Delta_{k+1:k+N_p}}{\sigma_{\Delta}},\;
x^n_k,\\
\frac{U^{0}_{{\mathrm{B}},k:k+N_p-1}}{P^{\max}_{\mathrm{B}}},\;
\frac{U^{0}_{{\mathrm{SC}},k:k+N_p-1}}{P^{\max}_{\mathrm{SC}}}
\biggr].
\label{eq:feature_vector}
\end{multline}
Here, $N_h$ is the history length, $N_p$ is the prediction horizon, and $\sigma_{\Delta}$ is the standard deviation of the disturbance signal used to normalize its magnitude. The disturbance history covers steps $k-N_h+1$ through $k$, and the preview covers steps $k+1$ through $k+N_p$; both windows are aligned so that step $k$ is the current control instant. The baseline command previews $U^{0}_{{\mathrm{B}},k:k+N_p-1}$ and $U^{0}_{{\mathrm{SC}},k:k+N_p-1}$ cover the same future steps $k$ through $k+N_p-1$ as the rollout commands in~\eqref{eq:projected_dpc_command}. The short horizon preview $\Delta_{k+1:k+N_p}$ is assumed to be available from a short term workload schedule over the prediction window. In the case studies, this preview is taken directly from the generated disturbance trajectory. Therefore, the present study isolates the control benefit of deterministic short horizon disturbance preview; forecasting errors are not considered.

The normalized HESS state is defined as
\begin{equation}
x^n_k
=
\begin{bmatrix}
P_{\mathrm{B}}[k]/P^{\max}_{\mathrm{B}} \\[3pt]
P_{\mathrm{SC}}[k]/P^{\max}_{\mathrm{SC}} \\[3pt]
\bigl(
\mathrm{SoC}_{\mathrm{B}}[k]
-
\mathrm{SoC}^{\mathrm{ref}}_{\mathrm{B}}
\bigr)
/\Delta\mathrm{SoC}_{\mathrm{B}} \\[3pt]
\bigl(
\mathrm{SoC}_{\mathrm{SC}}[k]
-
\mathrm{SoC}^{\mathrm{ref}}_{\mathrm{SC}}
\bigr)
/\Delta\mathrm{SoC}_{\mathrm{SC}} \\[3pt]
U_{\mathrm{B}}[k-1]/P^{\max}_{\mathrm{B}} \\[3pt]
U_{\mathrm{SC}}[k-1]/P^{\max}_{\mathrm{SC}}
\end{bmatrix}
\in \mathbb{R}^{6}.
\label{eq:normalized_hess_state}
\end{equation}
Here, $\Delta\mathrm{SoC}_{\mathrm{B}}$ and $\Delta\mathrm{SoC}_{\mathrm{SC}}$ denote the normalization spans used for the BESS and SC SoC deviations, respectively. The state vector collects the current device powers, SoC deviations from their operating references, and the commands applied at the previous sampling instant. Together with the disturbance history, preview, and baseline command preview, it provides the policy with the operating context required for residual command refinement.

The residual policy $\pi_{\Theta}$ is implemented as a fully connected feedforward neural network with three hidden layers. In the case studies, $N_h=64$ and $N_p=64$. From~\eqref{eq:feature_vector}, the input dimension is
\begin{equation}
N_h + N_p + 6 + N_p + N_p = 262,
\label{eq:input_dimension}
\end{equation}
and the output dimension is $2N_p=128$, corresponding to the two residual command sequences. GELU activation functions are used in the hidden layers. The final layer is initialized with zero weights and zero bias so that the untrained policy initially reproduces the rule-based baseline.

\subsection{Differentiable HESS Prediction Model}
\label{sec:differentiable_model}

The finite-horizon rollout uses the same storage dynamics as the HESS model. For $d\in\{\mathrm{B},\mathrm{SC}\}$, the device power follows the first-order discrete-time response obtained by ZOH discretization of~\eqref{eq:first_order_device_model}:
\begin{equation}
P_{d,j+1}
=
a_d P_{d,j}
+
b_d U_{d,j},
\label{eq:discrete_power_dynamics}
\end{equation}
with
\begin{equation}
a_d
=
e^{-T_s/\tau_d},
\qquad
b_d
=
1-a_d.
\label{eq:discrete_device_coefficients}
\end{equation}
Here, $j=0$ corresponds to the current control instant $k$, so $P_{d,0}=P_d[k]$ and $U_{d,0}$ determines the predicted power at the next sampling instant.

The SoC update uses the device power at the beginning of the sampling interval as a causal left endpoint approximation:
\begin{equation}
\mathrm{SoC}_{d,j+1}
=
\mathrm{SoC}_{d,j}
-
\frac{T_s}{3600\,E^{\mathrm{cap}}_d}
\cdot
\begin{cases}
\dfrac{P_{d,j}}{\eta^{\mathrm{dis}}_d},
&
P_{d,j}\geq 0,
\\[2mm]
P_{d,j}\eta^{\mathrm{ch}}_d,
&
P_{d,j}<0,
\end{cases}
\label{eq:soc_update}
\end{equation}
where $E^{\mathrm{cap}}_d$ is expressed in MWh, $P_{d,j}$ is expressed in MW, and $T_s$ is expressed in seconds. The factor $3600$ converts the sampling interval from seconds to hours. The charging and discharging efficiencies are denoted by $\eta^{\mathrm{ch}}_d$ and $\eta^{\mathrm{dis}}_d$, respectively. The piecewise efficiency model in~\eqref{eq:soc_update} is differentiable almost everywhere.

Collecting the device power and SoC into the state vector
\begin{equation}
x_j
=
\begin{bmatrix}
P_{\mathrm{B},j}\\
P_{\mathrm{SC},j}\\
\mathrm{SoC}_{\mathrm{B},j}\\
\mathrm{SoC}_{\mathrm{SC},j}
\end{bmatrix},
\label{eq:hess_state_vector}
\end{equation}
the prediction rollout is written compactly as
\begin{equation}
x_{j+1}
=
f_{\mathrm{HESS}}(x_j,U_j),
\qquad
j=0,\ldots,N_p-1,
\label{eq:hess_rollout}
\end{equation}
where $f_{\mathrm{HESS}}$ stacks the power dynamics~\eqref{eq:discrete_power_dynamics} and the SoC update~\eqref{eq:soc_update}. The rollout is implemented with automatic differentiation compatible operations, allowing gradients of the finite-horizon objective to be back-propagated to the policy parameters.

Within a prediction window initialized at control instant $k$, the previewed disturbance at prediction step $j+1$ corresponds to $\Delta[k+j+1]$. Because $U_j$ influences the device powers at step $j+1$, the predicted grid-side residual associated with the $j$-th control decision is
\begin{equation}
e_{g,j+1}
=
\Delta[k+j+1]
-
P_{\mathrm{B},j+1}
-
P_{\mathrm{SC},j+1}.
\label{eq:predicted_grid_residual}
\end{equation}
\subsection{Policy Training and Receding Horizon Deployment}
\label{sec:dpc_training}

The DPC policy is trained offline using disturbance windows sampled from the datacenter load trajectory. For each window, the policy generates a residual sequence, the differentiable HESS model predicts the closed-loop evolution, and the loss function penalizes grid-side residuals, command variation, excessive residual action, and operating limit violations.

The policy parameters are obtained from
\begin{equation}
\Theta^{\star}
=
\arg\min_{\Theta}
\frac{1}{nN_p}
\sum_{i=1}^{n}
\sum_{j=0}^{N_p-1}
\ell^i_{j+1}(\Theta),
\label{eq:training_objective}
\end{equation}
where $n$ is the batch size. The stage loss is
\begin{align}
\ell^i_{j+1}
={}&
Q_g
\left(
e^i_{g,j+1}
\right)^2
+
Q_t
\cdot
\mathbf{1}[j=N_p-1]
\cdot
\left(
e^i_{g,j+1}
\right)^2
\nonumber\\
&+
Q_{\Delta U}
\left\|
U^i_j-U^i_{j-1}
\right\|^2_2
+
Q_{\mathrm{soc}}
\ell^i_{\mathrm{soc},j+1}
+
Q_p
\ell^i_{p,j+1}
\nonumber\\
&+
Q_{\mathrm{ramp}}
\ell^i_{\mathrm{ramp},j+1}
+
Q_v
\ell^i_{\mathrm{viol},j+1}
+
Q_{\mathrm{res}}
\left\|
\delta U^i_j
\right\|^2_2.
\label{eq:stage_loss}
\end{align}
For $j=0$, $U^i_{-1}$ denotes the command applied immediately before the $i$-th training window; it is retained to ensure continuity of the command variation penalty at the window boundary and is included in the normalized HESS state~\eqref{eq:normalized_hess_state}. The first term penalizes the predicted grid-side residual at every step throughout the horizon. The second term is a terminal penalty active only at the final step $j=N_p-1$, where $\mathbf{1}[\cdot]$ denotes the indicator function. The command variation term encourages smooth control, and the residual regularization limits unnecessary deviations from the rule-based baseline.

The SoC displacement penalty is
\begin{equation}
\ell^i_{\mathrm{soc},j+1}
=
\left(
\mathrm{SoC}^{i}_{\mathrm{B},j+1}
-
\mathrm{SoC}^{\mathrm{ref}}_{\mathrm{B}}
\right)^2
+
\left(
\mathrm{SoC}^{i}_{\mathrm{SC},j+1}
-
\mathrm{SoC}^{\mathrm{ref}}_{\mathrm{SC}}
\right)^2.
\label{eq:soc_displacement_penalty}
\end{equation}

The power limit penalty is
\begin{equation}
\ell^i_{p,j+1}
=
\sum_{d\in\{\mathrm{B},\mathrm{SC}\}}
\left[
\max
\left(
|P^i_{d,j+1}|-P^{\max}_d,\,
0
\right)
\right]^2.
\label{eq:power_limit_penalty}
\end{equation}

The ramp rate penalty is
\begin{equation}
\ell^i_{\mathrm{ramp},j+1}
=
\sum_{d\in\{\mathrm{B},\mathrm{SC}\}}
\left[
\max
\left(
\left|
\frac{P^i_{d,j+1}-P^i_{d,j}}{T_s}
\right|
-
R^{\max}_d,\,
0
\right)
\right]^2.
\label{eq:ramp_limit_penalty}
\end{equation}

The SoC bound penalty is
\begin{equation}
\begin{aligned}
\ell^i_{\mathrm{viol},j+1}
=
\sum_{d\in\{\mathrm{B},\mathrm{SC}\}}
\Bigg\{
&
\left[
\max\left(
\mathrm{SoC}^{\min}_d
-
\mathrm{SoC}^{i}_{d,j+1},\,
0
\right)
\right]^2
\\
&+
\left[
\max\left(
\mathrm{SoC}^{i}_{d,j+1}
-
\mathrm{SoC}^{\max}_d,\,
0
\right)
\right]^2
\Bigg\}.
\end{aligned}
\label{eq:soc_bound_penalty}
\end{equation}
During offline training, the command sequence is projected through~\eqref{eq:projected_dpc_command}, while power output, ramp rate, and SoC operating requirements are incorporated through soft penalty terms. This structure allows the residual policy to improve smoothing performance without replacing the physically interpretable baseline allocation.

After training, the policy is deployed in a receding horizon fashion. At each time step $k$, the feature vector $\xi_k$ is assembled, the policy outputs a residual sequence, and only the first correction is applied:
\begin{equation}
\delta U_k
=
\begin{bmatrix}
\delta U_{\mathrm{B},0}\\
\delta U_{\mathrm{SC},0}
\end{bmatrix}.
\label{eq:first_residual_action}
\end{equation}
The candidate DPC command is
\begin{equation}
U^{\mathrm{DPC}}_d[k]
=
\Pi_d
\left(
U^{0}_d[k]
+
\delta U_d[k]
\right),
\qquad
d\in\{\mathrm{B},\mathrm{SC}\}.
\label{eq:candidate_dpc_command}
\end{equation}

A one-step safeguard uses the known disturbance preview $\Delta[k+1]$ to compare the predicted immediate residuals. Let $e^{\mathrm{DPC}}_g[k+1]$ and $e^{0}_g[k+1]$ denote the one-step predicted residuals under the candidate DPC command and the baseline command, respectively. Define
\begin{align}
U[k]
&=
\begin{bmatrix}
U_{\mathrm{B}}[k]\\
U_{\mathrm{SC}}[k]
\end{bmatrix},\quad
U^{\mathrm{DPC}}[k]
=
\begin{bmatrix}
U^{\mathrm{DPC}}_{\mathrm{B}}[k]\\
U^{\mathrm{DPC}}_{\mathrm{SC}}[k]
\end{bmatrix},\nonumber\\
U^{0}[k]
&=
\begin{bmatrix}
U^{0}_{\mathrm{B}}[k]\\
U^{0}_{\mathrm{SC}}[k]
\end{bmatrix}.
\label{eq:command_vectors}
\end{align}
The applied command is
\begin{equation}
U[k]
=
\begin{cases}
U^{\mathrm{DPC}}[k],
&
\left|
e^{\mathrm{DPC}}_g[k+1]
\right|
\leq
\left|
e^{0}_g[k+1]
\right|
+
\varepsilon_s,
\\[2mm]
U^{0}[k],
&
\text{otherwise}.
\end{cases}
\label{eq:dpc_safeguard}
\end{equation}

The safeguard is used as a conservative deployment time acceptance check for the first receding horizon correction. It does not replace the finite-horizon training objective and is not intended to provide a complete hard feasibility guarantee for power output, ramp rate, or SoC limits. Instead, it rejects candidate corrections that are predicted to produce a clearly inferior immediate residual relative to the baseline command. The tolerance $\varepsilon_s$ prevents the screening rule from discarding corrections whose one-step effect is nearly neutral. Thus, the rule-based baseline remains available as a reliable fallback command, whereas the residual DPC correction is applied whenever its immediate predicted effect is acceptable under the prescribed safeguard criterion.

\section{Simulation Studies}
\label{sec:case_studies}

\subsection{Simulation Setup}

The simulation studies are designed to assess both the local disturbance smoothing capability of the proposed HESS-DPC framework and its resulting impact on system frequency dynamics. The NPCC 140-bus system is simulated using the benchmark case distributed with the ANDES power system simulator~\cite{cui2020hybrid}. Synchronous generators are represented by the classical second-order GENCLS model in the NPCC benchmark. The analysis focuses on the resulting generator frequency responses to datacenter active power disturbances. Explicit excitation system and turbine governor dynamics are not included in this study. Therefore, the reported responses should be interpreted as forced frequency oscillations of the classical machine NPCC benchmark under the imposed datacenter disturbances. To introduce generator to generator diversity in the frequency response, the inertia coefficient of each GENCLS unit is scaled by a fixed random factor uniformly sampled from $[0.7,1.3]$, using the same random seed for all compared cases.

At the datacenter level, each datacenter is modeled as a 50-MW synthetic AI datacenter load following the aggregation model in Section~\ref{sec:system_modeling}. The load comprises three workload components, including one dominant training workload, one smaller training workload, and one fine-tuning workload. HESS-DPC is implemented locally at each datacenter. Its disturbance smoothing performance is therefore evaluated at the single datacenter level, while the system level frequency impact is assessed by injecting the compensated and uncompensated disturbance profiles from seven datacenters into the NPCC system. Each profile is applied at a separate load bus as a time-varying active power deviation, following the played in large load disturbance setup used in recent AI datacenter dynamic studies~\cite{ko2026wide,biswas2025evaluating,esig2026largeloaddisturbance}. The injection buses are selected as the largest load buses in the system, and the corresponding bus assignments are listed in Table~\ref{tab:dc_buses}. Different start times are used for the seven disturbance profiles to represent asynchronous workload operation, producing a peak aggregate disturbance of approximately 140~MW. The simulation horizon is 700~s with a sampling period of 0.01~s. The HESS device parameters and DPC training settings are listed in Table~\ref{tab:hess_parameters}.

\begin{figure}[t]
\centering
\includegraphics[width=\linewidth]{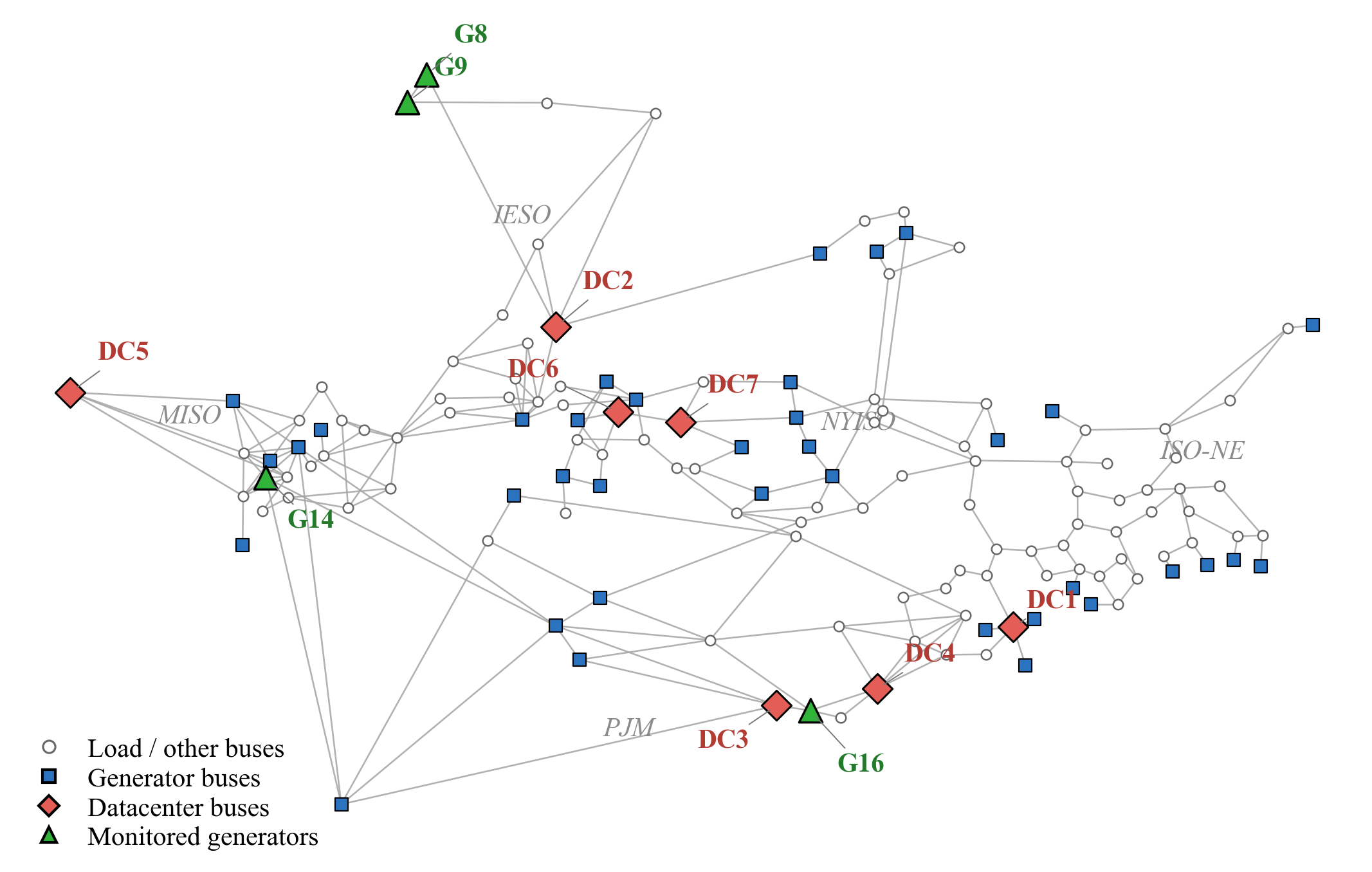}
\caption{NPCC 140-bus topology with datacenter disturbance injection buses and monitored generators highlighted.}
\label{fig:npcc_topology}
\end{figure}

In both training and closed-loop evaluation, the DPC policy uses a short-horizon disturbance preview over $N_pT_s=0.64$~s, consistent with the formulation in Section~\ref{sec:residual_dpc}. In the present case studies, this preview is taken from the generated workload trajectory so that the effect of predictive command refinement can be evaluated independently of disturbance-forecasting accuracy.

\begin{table}[t]
\centering
\renewcommand{\arraystretch}{1.08}
\setlength{\tabcolsep}{4.5pt}
\caption{Datacenter Disturbance Injection Buses in the NPCC 140-Bus System}
\label{tab:dc_buses}
\begin{tabular}{cccc}
\toprule
Datacenter & Bus & Base Load [MW] & Peak Deviation [MW] \\
\midrule
DC1 & 78  & 2000 & 17.1 \\
DC2 & 91  & 1650 & 18.5 \\
DC3 & 131 & 1160 & 16.5 \\
DC4 & 128 &  970 & 17.0 \\
DC5 & 120 &  946 & 18.4 \\
DC6 &  55 &  939 & 18.0 \\
DC7 &  53 &  923 & 16.6 \\
\bottomrule
\end{tabular}
\end{table}

\begin{figure}[t]
\centering
\includegraphics[width=\linewidth]{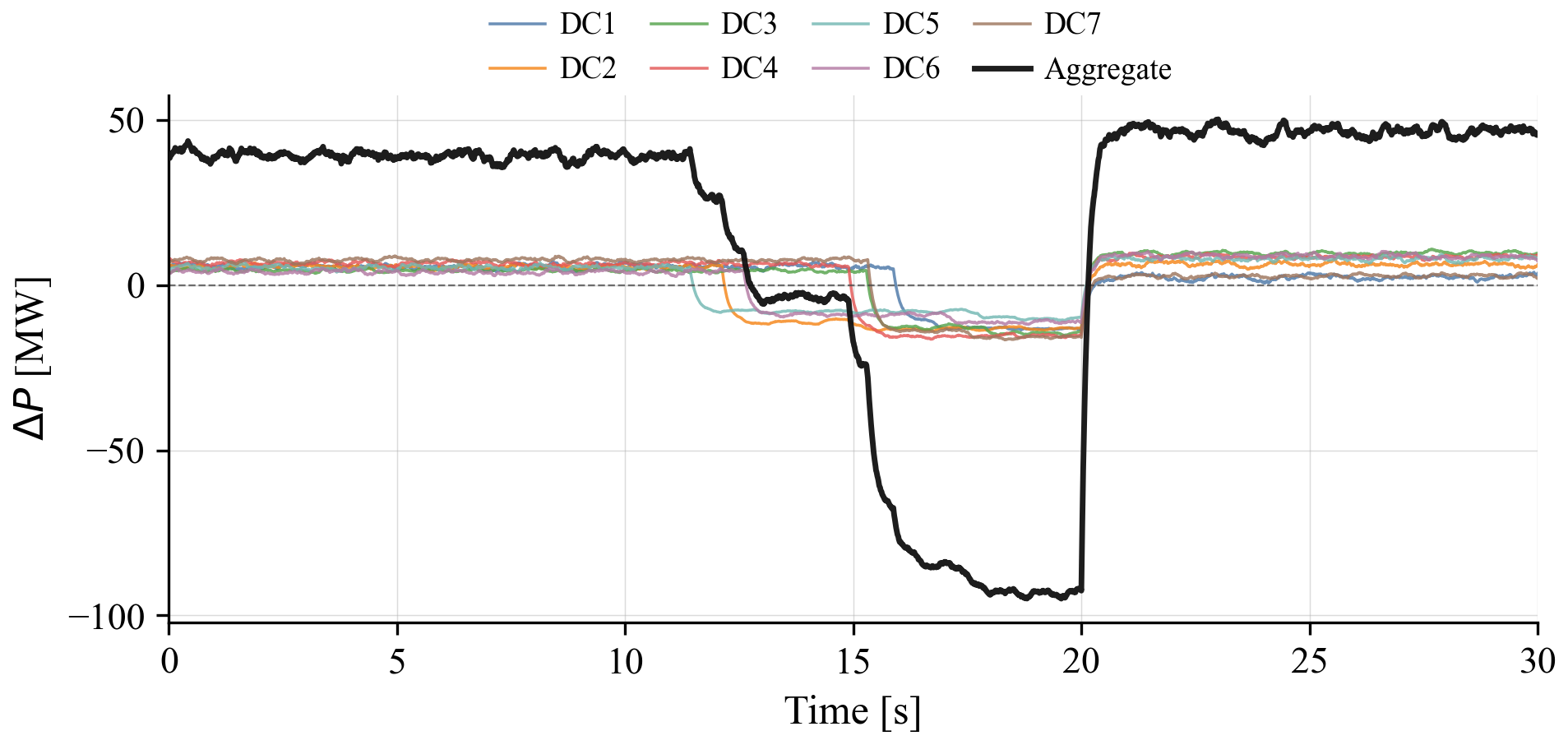}
\caption{Individual disturbance profiles of the seven datacenters and the corresponding aggregate deviation injected into the NPCC system.}
\label{fig:dc_disturbances}
\end{figure}

Fig.~\ref{fig:dc_disturbances} shows the individual and aggregate
disturbance profiles over a representative interval. Although the
individual profiles have comparable magnitudes, their superposition
produces a much larger aggregate deviation with both sustained plateaus
and sharp transitions, ranging from approximately $+50$~MW to near
$-90$~MW. This aggregate profile characterizes the uncompensated
system level disturbance used to assess the frequency impact in the
NPCC system.
\begin{table}[t]
\centering
\caption{Simulation Parameters for the HESS-DPC Case Studies}
\label{tab:hess_parameters}
\renewcommand{\arraystretch}{1.08}
\setlength{\tabcolsep}{4.5pt}
\begin{tabular}{lccc}
\toprule
Parameter & BESS & SC & Unit \\
\midrule
\multicolumn{4}{c}{\textit{Datacenter load parameters}} \\
\midrule
Nominal power of dominant training workload, $P^{\mathrm{tr}}_{L0}$
    & \multicolumn{2}{c}{50}    & MW   \\
Small training ratio, $\kappa_S$
    & \multicolumn{2}{c}{0.056} & --   \\
Fine-tuning ratio, $\kappa_F$
    & \multicolumn{2}{c}{0.056} & --   \\
\midrule
\multicolumn{4}{c}{\textit{HESS device parameters}} \\
\midrule
Power rating, $P_d^{\max}$
    & 30    & 15    & MW   \\
Energy capacity, $E_d^{\mathrm{cap}}$
    & 7   & 0.05  & MWh  \\
Ramp rate limit, $R_d^{\max}$
    & 50    & 100   & MW/s \\
Response time constant, $\tau_d$
    & 0.25  & 0.015 & s    \\
Initial state of charge, $\mathrm{SoC}_{d,0}$
    & 0.60  & 0.60  & --   \\
SoC lower bound, $\mathrm{SoC}_d^{\min}$
    & 0.05  & 0.05  & --   \\
SoC upper bound, $\mathrm{SoC}_d^{\max}$
    & 0.95  & 0.95  & --   \\
\midrule
\multicolumn{4}{c}{\textit{Baseline controller parameters}} \\
\midrule
Frequency split cutoff, $f_{\mathrm{split}}$
    & \multicolumn{2}{c}{0.5}  & Hz  \\
SC shaping cutoff, $f_{\mathrm{shape}}$
    & \multicolumn{2}{c}{8.0}  & Hz  \\
\midrule
\multicolumn{4}{c}{\textit{DPC policy parameters}} \\
\midrule
Prediction horizon, $N_p$
    & \multicolumn{2}{c}{64}    & steps \\
History window, $N_h$
    & \multicolumn{2}{c}{64}    & steps \\
Safety tolerance, $\varepsilon_s$
    & \multicolumn{2}{c}{0.05}  & MW    \\
Training epochs
    & \multicolumn{2}{c}{800}   & --    \\
Training/validation split
    & \multicolumn{2}{c}{85/15} & \%    \\
\bottomrule
\end{tabular}
\end{table}

\subsection{HESS Power Smoothing}
\begin{figure}[t]
\centering
\includegraphics[width=\linewidth]{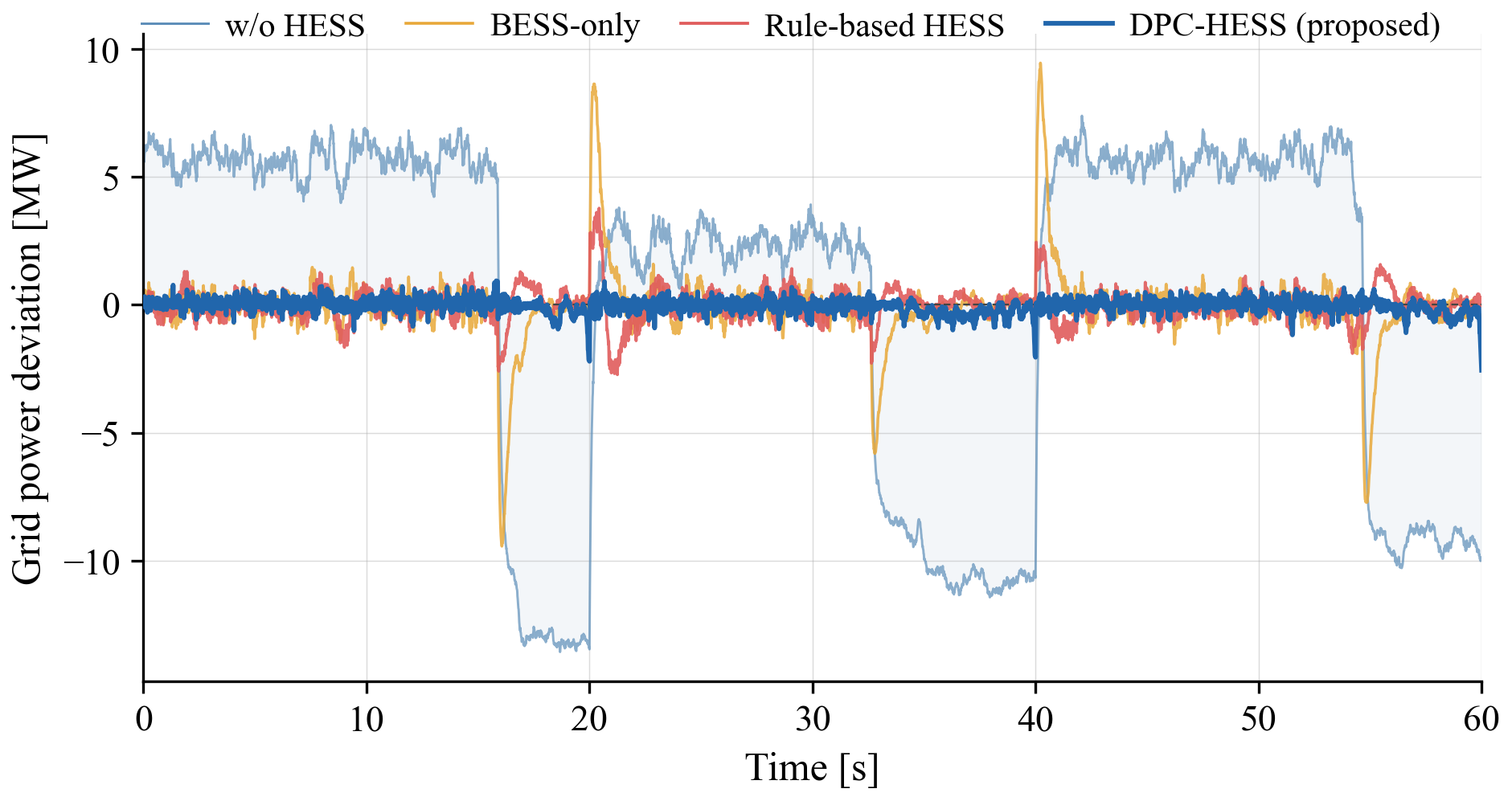}
\caption{Grid side residual power deviation under the four smoothing configurations.}
\label{fig:grid_deviation}
\end{figure}

Four configurations are compared: (A)~no HESS, (B)~BESS only, (C)~rule-based HESS, and (D)~the proposed HESS-DPC. The grid-side residual power deviation $P_{\mathrm{grid}}^{\mathrm{dev}}$ under these four configurations is shown in Fig.~\ref{fig:grid_deviation}. Without HESS, the residual closely follows the workload-induced fluctuation of the datacenter load. The BESS-only configuration reduces the slowly varying component, but large residual spikes remain around the major workload transition intervals because the BESS ramp rate limit and response time constant prevent it from tracking rapid power changes.

The rule-based HESS substantially reduces these transient deviations by
assigning the fast disturbance component to the SC and the remaining
component to the BESS. Nevertheless, residual deviations persist near
workload transition intervals, as the fixed frequency decomposition
does not incorporate preview information about the upcoming disturbance
profile.

The proposed HESS-DPC achieves the smallest residual among all four
cases. The residual remains within a small band around zero during
steady fluctuation intervals and is strongly attenuated at workload
transition intervals---precisely the operating condition where fixed
frequency decomposition is most limited due to the absence of disturbance
preview. These results confirm that a predictive residual correction over
a short horizon of $N_pT_s{=}0.64$~s substantially reduces the tracking
errors at workload transitions while preserving the structured BESS/SC
allocation.

\begin{figure}[t]
\centering
\includegraphics[width=\linewidth]{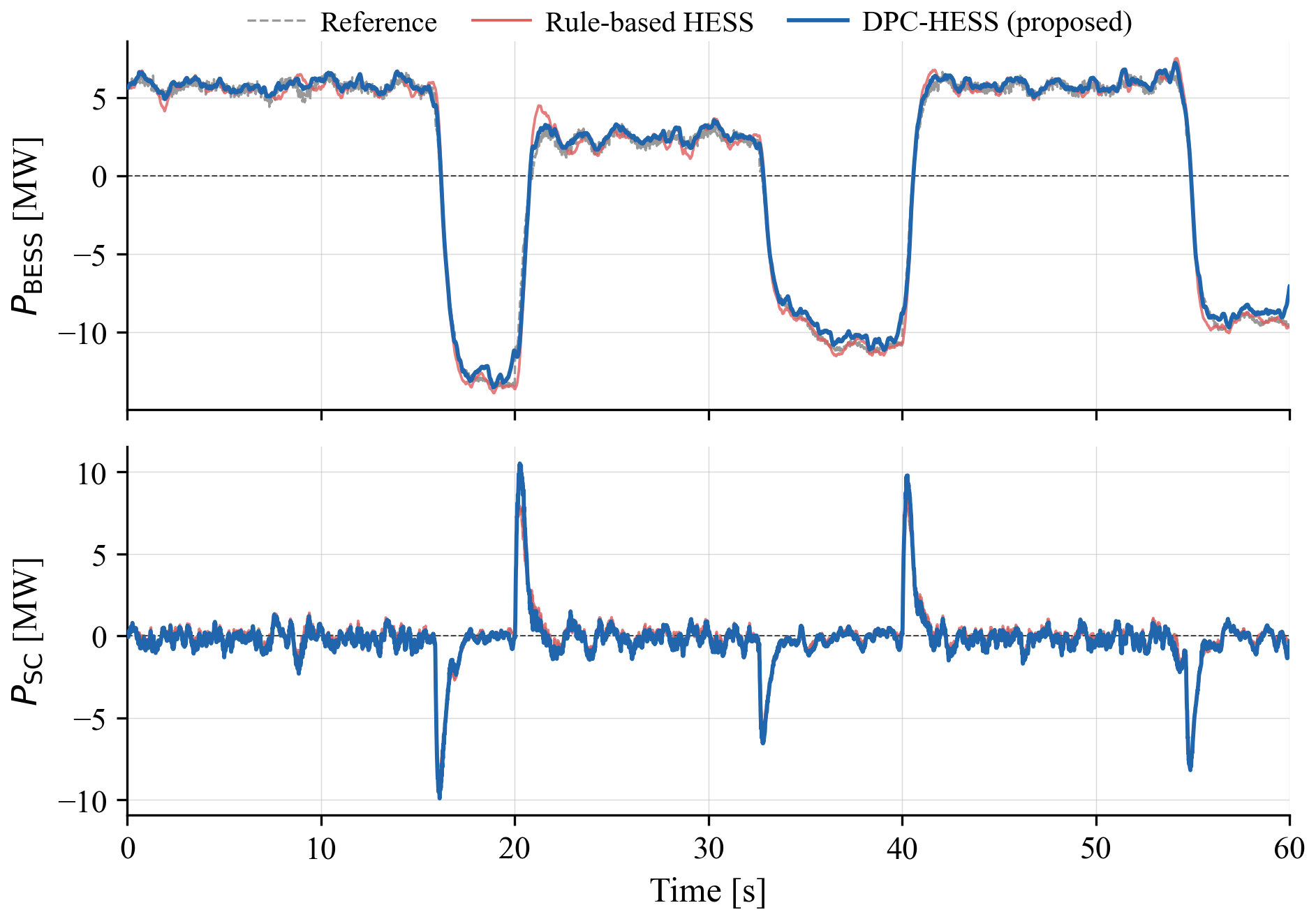}
\caption{BESS and SC power outputs under the rule-based HESS and proposed HESS-DPC controllers, together with their reference signals.}
\label{fig:hess_components}
\end{figure}

Fig.~\ref{fig:hess_components} shows the BESS and SC power outputs under
the rule-based HESS and HESS-DPC configurations, together with their
reference signals. The BESS primarily tracks the low frequency
energy-dominant component of the disturbance, while the SC responds to
fast transitions and high-frequency variations. Under both controllers,
the device outputs closely follow their references, confirming that the
frequency-based allocation is consistent with the physical
characteristics of the two devices.

Under HESS-DPC, the residual corrections are modest in magnitude but
concentrated at workload transition intervals. The SC correction
supplements the rule-based shaping near workload transition intervals,
while the BESS correction reduces slower tracking errors. The resulting
power trajectories remain within the rated ranges in
Table~\ref{tab:hess_parameters} throughout the simulation.

\begin{figure}[t]
\centering
\includegraphics[width=\linewidth]{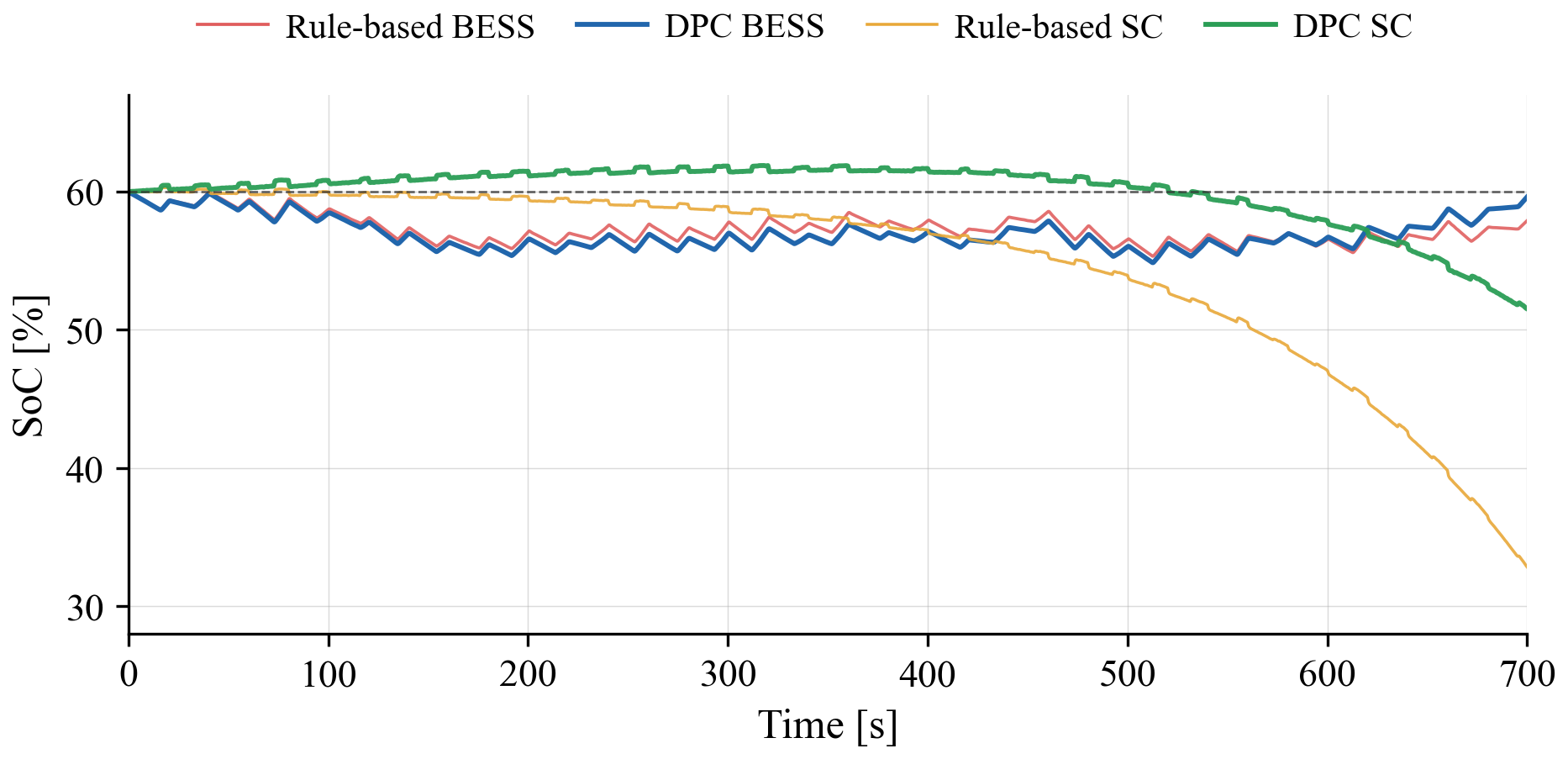}
\caption{SoC trajectories of the BESS and SC under the rule-based HESS and proposed HESS-DPC controllers over the 700 s simulation horizon.}
\label{fig:soc}
\end{figure}

The SoC trajectories over the full 700 s horizon are shown in
Fig.~\ref{fig:soc}. Both devices are initialized at
$\mathrm{SoC}_0 = 0.60$. The BESS SoC remains close to its initial value
under both controllers, consistent with the zero-mean property of
$\Delta[n]$ established in Section~\ref{sec:system_modeling}. A more pronounced difference is observed
in the SC SoC. Under the rule-based controller, the SC SoC drifts
progressively downward during the second half of the simulation,
indicating that the fixed rule-based allocation accumulates a net energy
imbalance over an extended operating horizon. Under HESS-DPC, the SC SoC
remains substantially closer to its initial level throughout the
simulation, confirming that the finite-horizon policy improves SC energy
sustainability over the extended horizon without a dedicated SoC
restoration scheme.

The one-step safeguard in~\eqref{eq:dpc_safeguard} governs which DPC
corrections are applied at each time step. A smaller tolerance
\(\varepsilon_{\mathrm{s}}\) makes the safeguard more conservative,
accepting the DPC correction only when its predicted one-step residual
is no larger than the baseline residual within the prescribed tolerance.
A larger \(\varepsilon_{\mathrm{s}}\) admits more corrections, including
steps for which the predicted immediate advantage over the baseline is
smaller.

Table~\ref{tab:eps_sensitivity} reports the DPC acceptance rate, RMS
grid-side residual, RMS reduction relative to the rule-based baseline,
peak-to-peak residual, and minimum SC SoC over the 700 s horizon for
five values of \(\varepsilon_{\mathrm{s}}\). The rule-based HESS
baseline achieves an RMS residual of 1.36~MW and a peak-to-peak
residual of 11.45~MW. At \(\varepsilon_{\mathrm{s}}=0\), the safeguard
accepts 45.8\% of DPC corrections and reduces the RMS residual to
0.300~MW, corresponding to a 77.9\% reduction relative to the
rule-based baseline. As \(\varepsilon_{\mathrm{s}}\) increases from
0 to 0.20~MW, the acceptance rate rises from 45.8\% to 60.6\%, while
the RMS reduction decreases from 77.9\% to 72.6\%. This trend indicates
that the additional corrections admitted under more relaxed thresholds
provide weaker immediate improvement on average, leading to moderately
degraded residual power metrics. Across the tested range, the RMS reduction remains above 72\% and the minimum SC SoC stays above 50\%, showing that the safeguard maintains strong smoothing performance over a practical range of tolerance values. In the remaining case studies, $\varepsilon_{\mathrm{s}}=0.05$~MW is used as a balanced setting between correction acceptance and residual quality.

\begin{table}[t]
\centering
\renewcommand{\arraystretch}{1.08}
\setlength{\tabcolsep}{3.2pt}
\caption{Sensitivity of HESS-DPC Performance to
\(\varepsilon_{\mathrm{s}}\)}
\label{tab:eps_sensitivity}
\begin{tabular}{cccccc}
\toprule
\(\varepsilon_{\mathrm{s}}\)
& Accept.
& RMS
& RMS
& Peak-to-peak
& Min SC \\
{[MW]}
& rate {[\%]}
& residual {[MW]}
& reduction {[\%]}
& residual {[MW]}
& SoC {[\%]} \\
\midrule
Rule based & --   & 1.36  & --   & 11.45 & --    \\
0.00       & 45.8 & 0.300 & 77.9 & 4.91  & 50.91 \\
0.01       & 46.9 & 0.303 & 77.7 & 5.00  & 51.06 \\
0.05       & 50.7 & 0.316 & 76.8 & 5.05  & 51.52 \\
0.10       & 54.6 & 0.336 & 75.3 & 5.32  & 52.03 \\
0.20       & 60.6 & 0.373 & 72.6 & 5.71  & 53.11 \\
\bottomrule
\end{tabular}
\end{table}

\subsection{Frequency Impact on the NPCC 140-Bus System}

\begin{figure}[t]
\centering
\includegraphics[width=\linewidth]{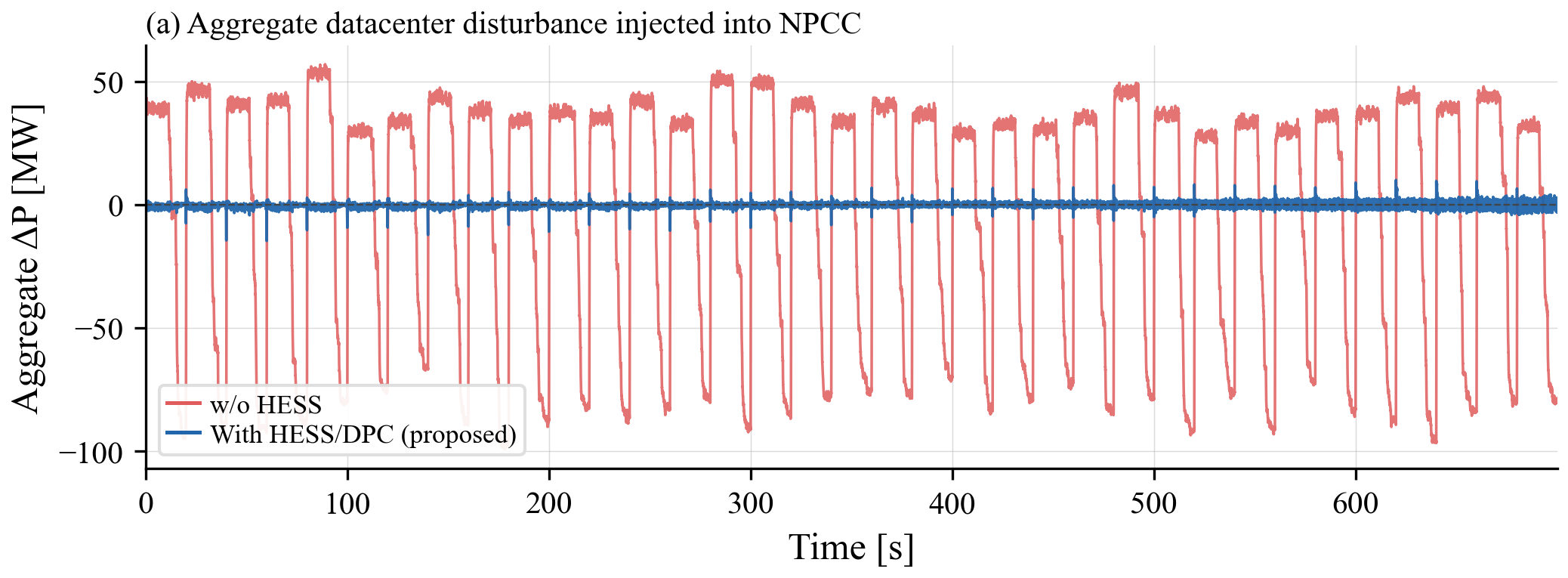}
\vspace{0.3em}
\includegraphics[width=\linewidth]{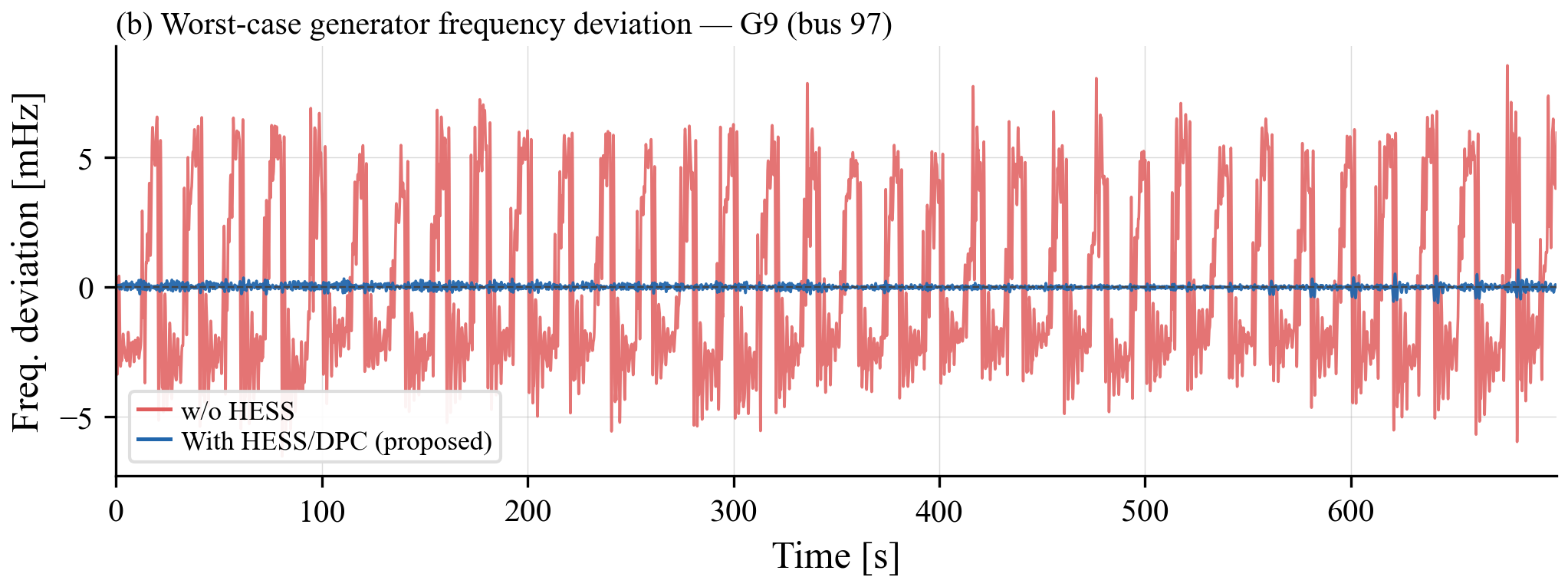}
\vspace{0.3em}
\includegraphics[width=\linewidth]{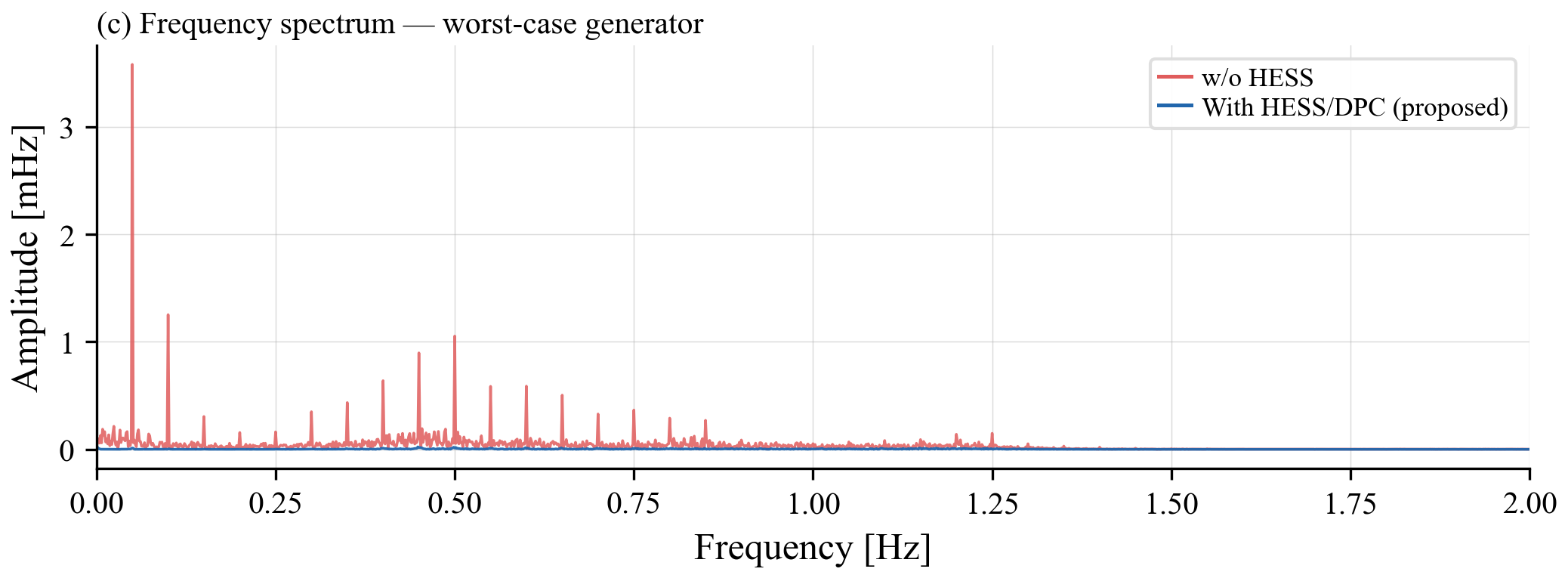}
\caption{Datacenter disturbance smoothing impact on the NPCC 140-bus system: aggregate disturbance, generator frequency response, and frequency spectrum.}
\label{fig:npcc_response}
\end{figure}

Fig.~\ref{fig:npcc_response}(a) shows the aggregate power deviation from
the seven datacenters over the full simulation horizon. Without HESS,
the aggregate disturbance exhibits sustained large-amplitude fluctuations
whose cycle to cycle amplitudes vary due to the stochastic workload
profiles and asynchronous timing offsets among the facilities. With the
proposed HESS-DPC applied at each datacenter, the aggregate residual is
suppressed to a narrow band around zero, showing that local compensation at each datacenter remains effective once the seven profiles are combined.

Fig.~\ref{fig:npcc_response}(b) shows the frequency deviation of
generator G9 at bus~97, identified as the most affected generator under
this disturbance scenario. Without HESS, G9 exhibits sustained
oscillations throughout the simulation horizon with peak-to-peak
deviations reaching 15.1~mHz. These oscillations are consistent with
the structured and periodic nature of the aggregate datacenter
disturbance. This behavior indicates that the uncompensated disturbance
acts as a persistent forcing input to the bulk system. With the proposed
HESS-DPC, the frequency deviation is strongly attenuated and remains
close to zero, substantially reducing the sustained oscillatory
excitation in the bulk system.

The frequency spectrum of the G9 deviation is shown in
Fig.~\ref{fig:npcc_response}(c). Without HESS, the spectrum exhibits a
dominant low frequency peak at the fundamental workload cycle frequency
and several harmonic components, each of which may interact with
electromechanical modes of the power system. With HESS-DPC compensation,
the spectral amplitude is reduced to near the noise floor across the
range below 2~Hz.

\begin{figure}[t]
\centering
\includegraphics[width=\linewidth]{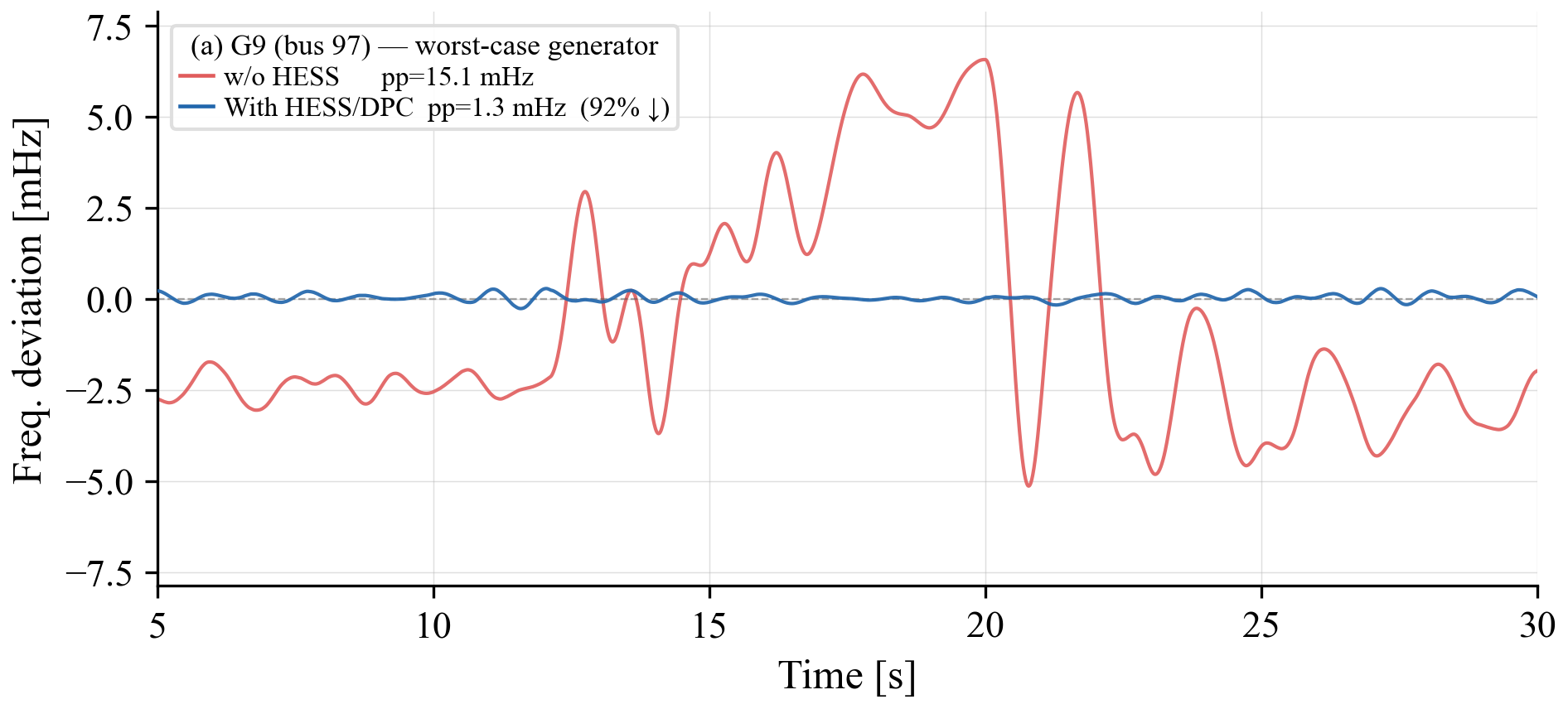}
\caption{Zoomed view of the G9/bus~97 frequency deviation. HESS-DPC reduces the peak-to-peak value from 15.1~mHz to 1.3~mHz.}
\label{fig:g9_zoom}
\end{figure}

\begin{figure}[t]
\centering
\includegraphics[width=\linewidth]{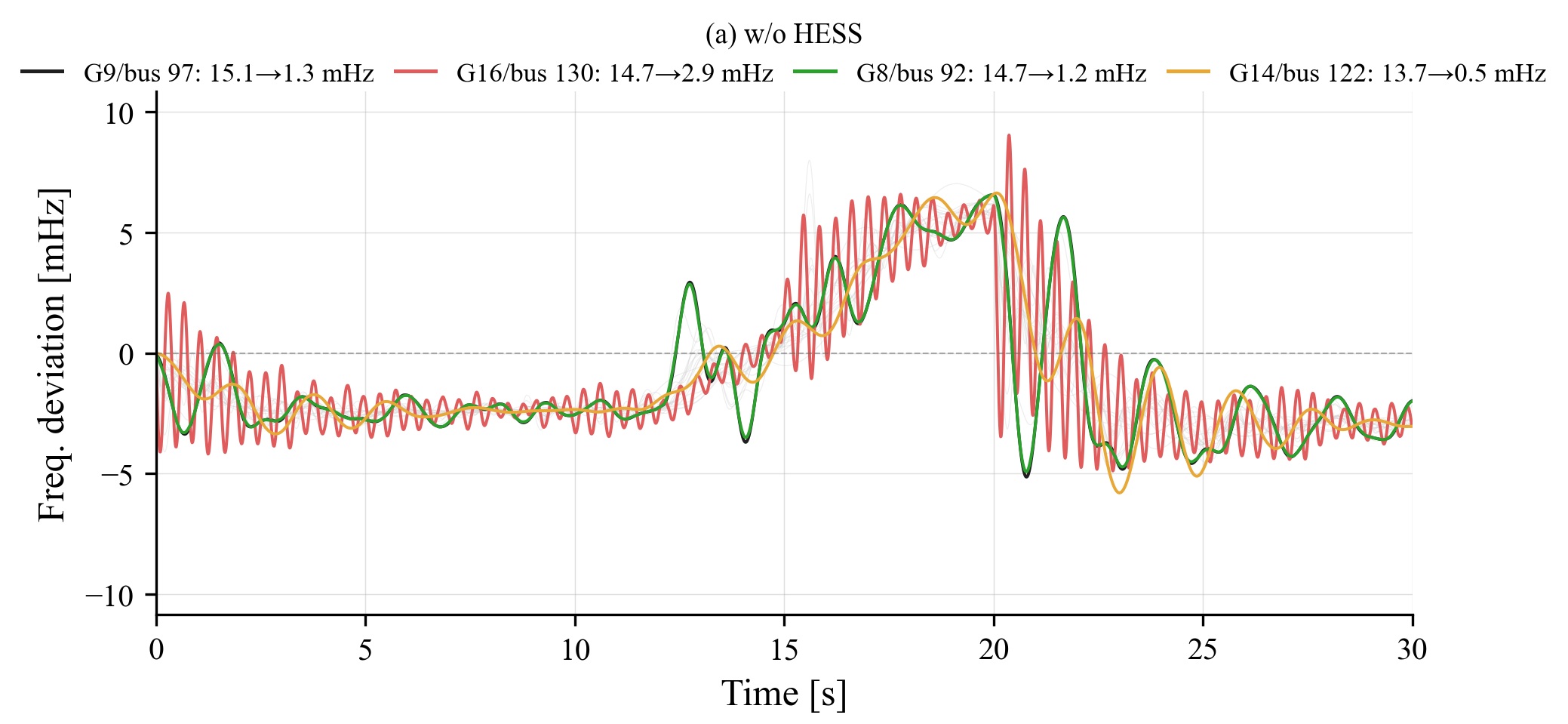}
\vspace{0.3em}
\includegraphics[width=\linewidth]{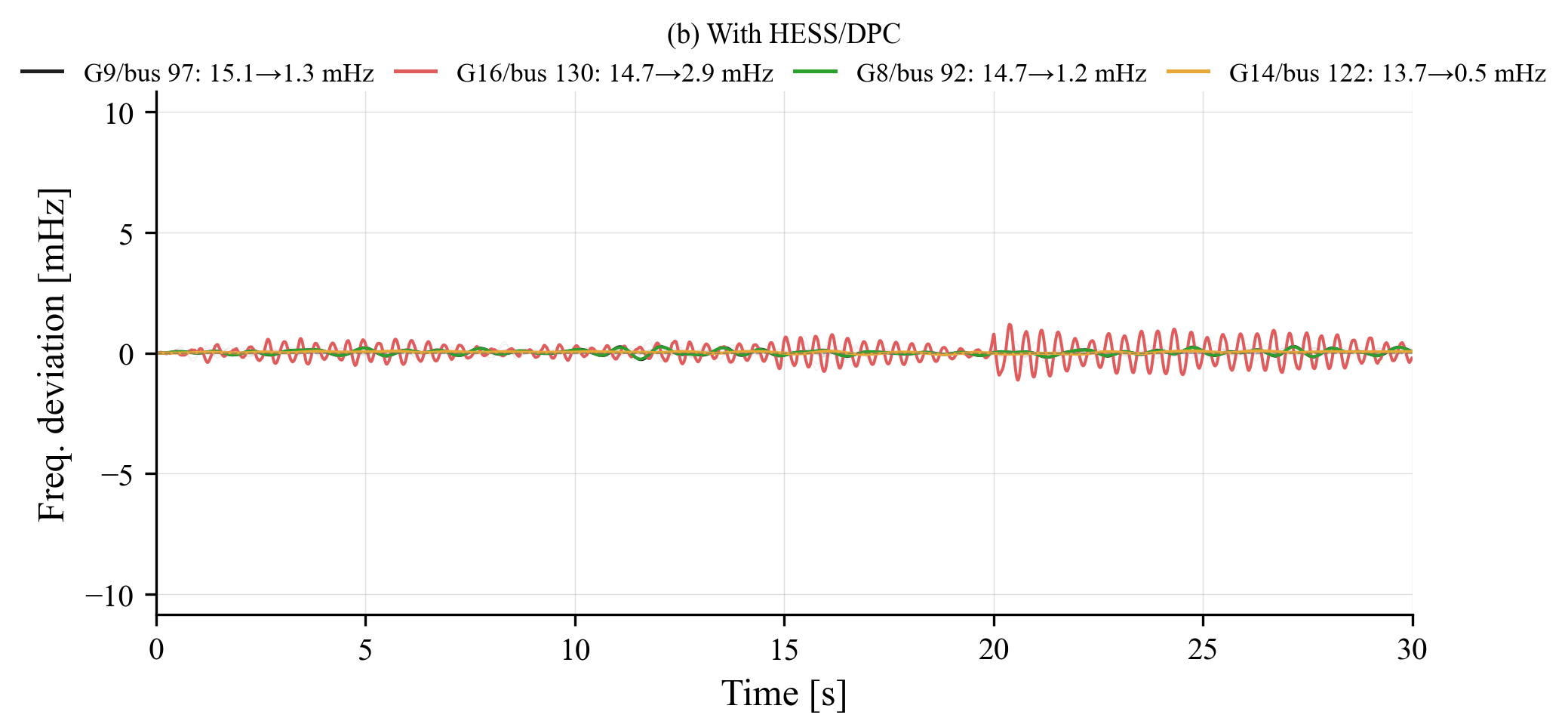}
\caption{Frequency deviations of the four most affected generators under uncompensated and HESS-DPC-compensated datacenter disturbances.}
\label{fig:multi_generator}
\end{figure}

To quantify the mitigation effect, Fig.~\ref{fig:g9_zoom} provides a
zoomed view of the G9 frequency deviation over a representative 25 s
interval. Without HESS, the peak-to-peak deviation reaches 15.1~mHz.
With the proposed HESS-DPC, this value is reduced to 1.3~mHz,
corresponding to a 91.4\% reduction. Fig.~\ref{fig:multi_generator}
extends this comparison to the four most affected generators, with the
corresponding peak-to-peak values summarized in
Table~\ref{tab:frequency_reduction}. Without HESS, all four generators
exhibit sustained oscillatory responses with peak-to-peak deviations in
the range of 13--16~mHz, sharing the periodicity of the aggregate
forcing disturbance but differing in phase and waveform shape due to
their distinct modal participation and network coupling. With the
proposed HESS-DPC, the peak-to-peak deviation of each generator is
reduced by more than 80\%. A small residual oscillation remains at G16,
which is more strongly coupled to the dominant system mode excited by
the disturbance, but its compensated deviation remains at the mHz level
and is much smaller than the uncompensated response. These results
demonstrate that local power smoothing at the datacenter level
effectively reduces frequency deviations across the NPCC 140-bus system.

\begin{table}[t]
\centering
\renewcommand{\arraystretch}{1.08}
\caption{Peak-to-Peak Frequency Deviation Reduction of Monitored Generators}
\label{tab:frequency_reduction}
\begin{tabular}{lccc}
\toprule
Generator & Without HESS [mHz] & HESS-DPC [mHz] & Reduction [\%] \\
\midrule
G9 / bus 97   & 15.1 & 1.3 & 91.4 \\
G16 / bus 130 & 14.7 & 2.9 & 80.3 \\
G8 / bus 92   & 14.7 & 1.2 & 91.8 \\
G14 / bus 122 & 13.7 & 0.5 & 96.4 \\
\bottomrule
\end{tabular}
\end{table}

The classical GENCLS benchmark excludes governor and excitation dynamics,
which provide additional frequency damping in practice. To examine whether
this simplification affects the estimated mitigation benefit, the same
datacenter disturbance setting is repeated using the full dynamic NPCC
model. The full dynamic model retains 21 GENCLS units and additionally
includes 27 GENROU machines, 29 TGOV1 turbine governors, and 24 IEEEX1
excitation systems from the original NPCC dynamic data; the smaller
frequency response it produces relative to the classical benchmark reflects
this additional damping. Cases without HESS-DPC and with HESS-DPC are
evaluated for both model representations.

\begin{figure}[t]
\centering
\begin{subfigure}{0.95\linewidth}
    \centering
    \includegraphics[width=\linewidth]{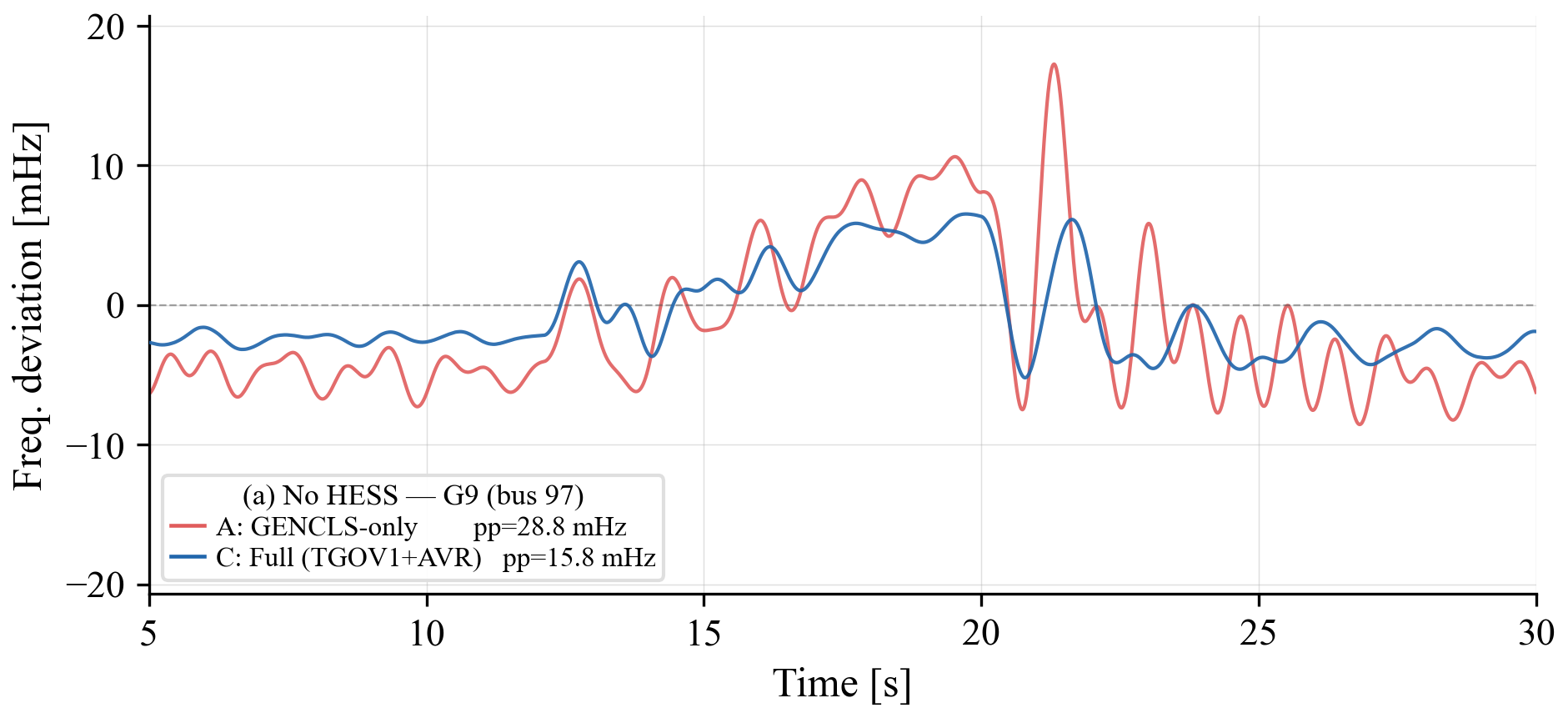}
    \caption{Without HESS-DPC.}
    \label{fig:model_fidelity_no_hess}
\end{subfigure}

\vspace{0.4em}

\begin{subfigure}{0.95\linewidth}
    \centering
    \includegraphics[width=\linewidth]{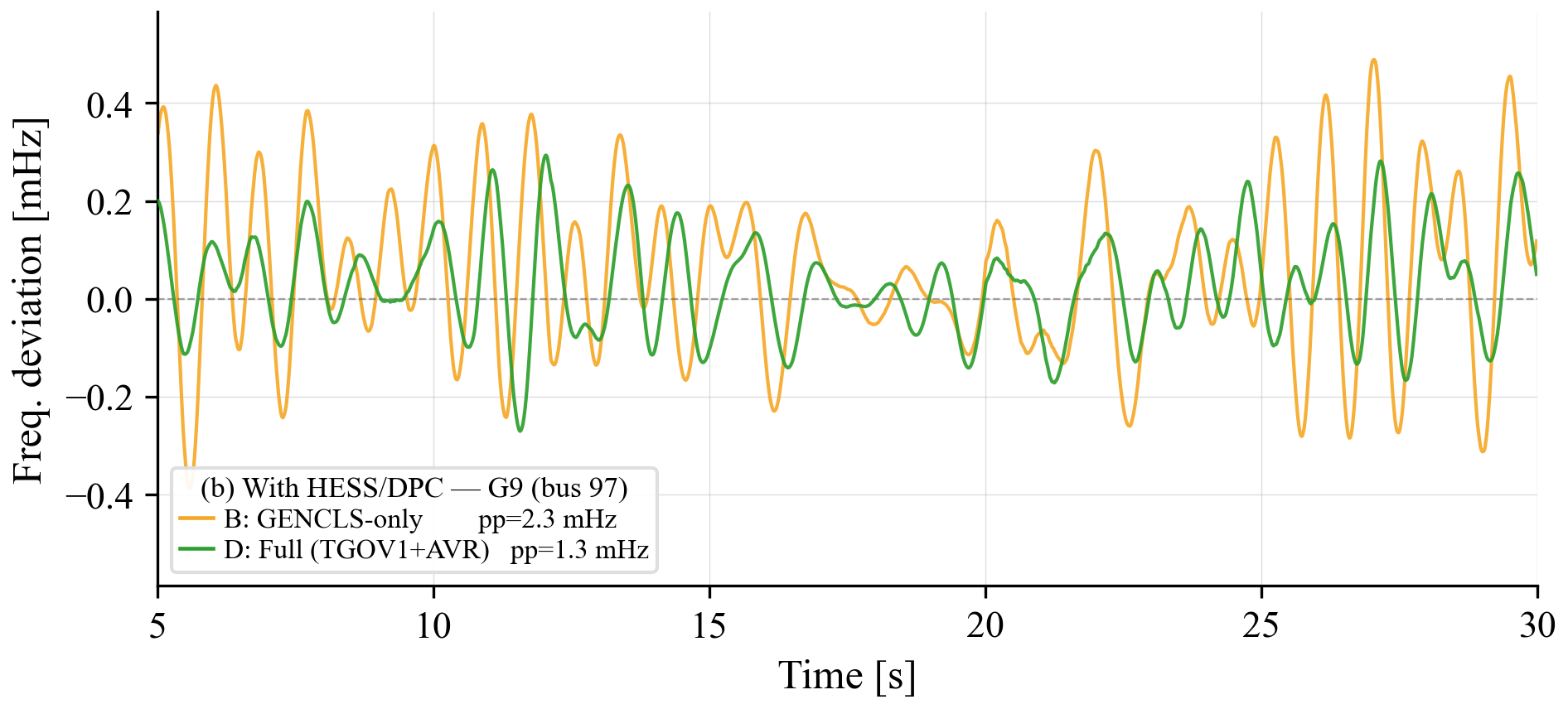}
    \caption{With HESS-DPC.}
    \label{fig:model_fidelity_with_hess}
\end{subfigure}

\caption{Effect of generator dynamic representation on the G9/bus 97 frequency response:
(a) without HESS-DPC and (b) with HESS-DPC.}
\label{fig:model_fidelity}
\end{figure}

Figure~\ref{fig:model_fidelity} compares the frequency response of the
common reference generator G9 at bus 97. Without HESS-DPC, the
peak-to-peak deviation is 28.8~mHz in the classical machine NPCC
benchmark and 15.8~mHz in the full dynamic model. With HESS-DPC, the
corresponding values are reduced to 2.3~mHz and 1.3~mHz, respectively.
The same trend is observed at the system level. Based on the worst
generator response in each case, HESS-DPC reduces the peak-to-peak
deviation from 28.80~mHz to 3.07~mHz in the classical machine NPCC
benchmark, corresponding to an 89.3\% reduction. In the full dynamic
model, the worst generator deviation decreases from 15.83~mHz to
2.94~mHz, corresponding to an 81.4\% reduction. Thus, the full dynamic
model confirms that HESS-DPC strongly suppresses the frequency response
caused by datacenter power fluctuations. At the same time, the
classical machine NPCC benchmark overestimates the relative mitigation
benefit by 7.9 percentage points.

The case studies verify the proposed source-side mitigation path:
HESS-DPC suppresses the active-power disturbance at the datacenter
point of interconnection under time-varying workload conditions, with
the residual DPC correction contributing most at workload transitions
where fixed frequency decomposition falls short. The reduced disturbance
injected into the NPCC 140-bus system results in substantially lower
generator frequency deviations and spectral excitation across the system.
The full dynamic model comparison shows that this mitigation benefit
persists when governor and excitation dynamics are represented, so the
result is not an artifact of the simplified GENCLS benchmark.

\section{Conclusion}
\label{sec:conclusion}

This paper presented a source-side active power smoothing framework in
which a frequency-decomposition hybrid energy storage system provides a
structured rule-based baseline and a residual differentiable predictive
control policy refines the baseline commands using a short-horizon
disturbance preview. In NPCC 140-bus simulations with seven 50~MW AI
datacenters, HESS-DPC reduced grid-side residual power deviations at the
point of interconnection and maintained SC state-of-charge balance over
extended operation; generator peak-to-peak frequency deviations were
reduced by more than 80\% across all monitored generators, with the
worst-case generator (G9 at bus~97) reduced from 15.1~mHz to 1.3~mHz.
This gain stems from the complementary roles of the two layers: the
rule-based frequency decomposition handles steady-state allocation,
while the residual DPC correction applies most of its action at workload
transition intervals, where the fixed baseline cannot anticipate the
approaching disturbance change. Repeating the evaluation with a full
dynamic NPCC model that includes governor and excitation systems yields
the same more-than-80\% reduction, showing that the result does not
depend on the simplified GENCLS representation. These results
demonstrate that source-side smoothing at the datacenter point of
interconnection, reinforced by a short-horizon learned correction, can
substantially limit the impact of AI workload fluctuations on bulk
system frequency.

\ifCLASSOPTIONcaptionsoff
  \newpage
\fi

\bibliographystyle{IEEEtran}
\bibliography{REF}

\end{document}